\documentclass[12pt]{article}

\usepackage{setspace}
\usepackage[square,numbers,compress]{natbib}
\usepackage[utf8]{inputenc}
\usepackage{amsfonts}
\usepackage{amssymb}
\usepackage{multirow}
\usepackage{amsmath}
\usepackage{titling}
\usepackage{mathrsfs}
\usepackage{lscape}
\usepackage[margin=1.5in]{geometry}
\usepackage{caption}
\usepackage{graphicx}
\usepackage{subcaption}
\usepackage{setspace}
\usepackage{multibib}
\usepackage{lipsum}
\usepackage{tcolorbox}
\usepackage{xcolor}
\usepackage[section]{placeins}
\usepackage{chngcntr}
\usepackage{newfloat}
\usepackage{booktabs}
\usepackage{tabularx}
\usepackage{ltablex}
\usepackage{multicol}
\usepackage{authblk}
\usepackage{lineno}
\usepackage{har2nat}
\usepackage{setspace}
\usepackage[onelanguage,ruled,vlined,linesnumbered,resetcount]{algorithm2e}
\usepackage[pdftex,hidelinks]{hyperref}
 
\usepackage[section]{placeins}
\usepackage{authblk}

\title{Endogenous Labour Flow Networks}

\author[1,*]{Kathyrn R. Fair}
\author[1]{Omar A. Guerrero}

\affil[1]{The Alan Turing Institute, London}
\affil[*]{kfair@turing.ac.uk}

\date{}

\begin{document}

\maketitle





\begin{abstract}
In the last decade, the study of labour dynamics has led to the introduction of labour flow networks (LFNs) as a way to conceptualise job-to-job transitions, and to the development of mathematical models to explore the dynamics of these networked flows. To date, LFN models have relied upon an assumption of static network structure. However, as recent events (increasing automation in the workplace, the COVID-19 pandemic, a surge in the demand for programming skills, etc.) have shown, we are experiencing drastic shifts to the job landscape that are altering the ways individuals navigate the labour market. Here we develop a novel model that emerges LFNs from agent-level behaviour, removing the necessity of assuming that future job-to-job flows will be along the same paths where they have been historically observed. This model, informed by microdata for the United Kingdom, generates empirical LFNs with a high level of accuracy. We use the model to explore how shocks impacting the underlying distributions of jobs and wages alter the topology of the LFN. This framework represents a crucial step towards the development of models that can answer questions about the future of work in an ever-changing world.
\end{abstract}

\textbf{Teaser: Emerging complex labour mobility patterns from the bottom up} 

\section*{Introduction}

Labour markets are highly heterogeneous complex systems that shape the economy of every country in the world.
Recently, technological changes such as the automation of work and worldwide shocks like the COVID-19 pandemic have produced structural changes that are reshaping labour mobility in new ways.
For example, remote working has enabled new employment opportunities to people who, previously, may not have applied or being considered for those positions due to geographical constraints.
A decade ago, a body of literature that employs network-science methods emerged and grew under the umbrella of \textit{labour flow networks} (LFNs) \cite{guerrero_employment_2013, collet_old_2013, schmutte_free_2014, guerrero_firmtofirm_2015, guerrero_understanding_2016, axtell_frictional_2019, park_global_2019, lopez_network_2020, delrio-chanona_occupational_2021, moro_universal_2021, applegate_job_2022, alabdulkareem_unpacking_2018}.
Much of these works, however, can only address short-term dynamics, as they operate under the assumption of a constancy in labour market conditions (no structural changes).
Thus, new modelling frameworks that can explain the endogenous formation of LFNs from agent-level behaviour (as opposed to historical data on labour flows) are necessary.
This paper develops one such framework by combining insights from the LFN literature with well-accepted microfoundations of economic labour market models.
We calibrate our model to the UK labour market and conduct a systematic analysis of how the UK LFN responds to changes in the distribution of job positions and wages; two factors that are susceptible to global shocks, free trade agreements, technological change, global supply chain disruptions, etc.
We find that LFN structure is more sensitive to changes in the job distribution than in the wage distribution, and that the extent of impacts depends not only on how many positions are affected but on which industries these positions belong to.
To the extent of our knowledge, this is the first model that is able to reproduce the topology of LFNs with a high fidelity (i.e. is not limited to reproducing stylised facts like degree distributions) through agent-level behaviour and, thus, it represents a milestone towards creating models that can address a broad class of dynamical problems about labour and, more generally, the future of work.

The rest of the paper is organised in the following way.
First, we review recent research concerning labour flow networks to situate our own work within this literature.
Then, we present the main results from our modelling exercise. 
Following this, we provide conclusions and discuss avenues for advancement.
Finally, we detail our modelling framework, including the data used to inform the model and the methodology used for calibration.

\subsection*{On the study of labour flow networks}

In recent years, network science methods have helped improving our understanding of labour dynamics at a highly desegregated level.
This strand of research can be encompassed under the umbrella of LFNs.
By using data on employment histories recorded through surveys, administrative databases, and online recruitment platforms, LFN studies characterise the labour market as a complex network where nodes represent jobs or sets of jobs (e.g., firms, industries, occupations, etc.) and edges indicate observed flows between them.
These networks reveal structural properties of the labour market with potential implications in their dynamics.
Importantly, this literature differentiates itself from the sociological \cite{granovetter_strength_1973} and economic \cite{calvo-armengol_job_2004} traditions focusing on the diffusion of vacancy information through social networks (see \cite{marsden_social_2001,beaman_social_2016} for comprehensive surveys).\footnote{The LFN literature differentiates itself from relatedness studies \cite{alabdulkareem_unpacking_2018, farjoun_independent_1998, boschma_how_2009,  neffke_value_2019} that adopt metrics of economic complexity \cite{hidalgo_product_2007} to analyse the co-occurrence of skills in occupations and job places.}
Instead LFN studies view realised labour flows as a source of information to infer structural properties of the labour market.
For example, a core-periphery structure in a firm-based LFN reveals highly compartmentalised dynamics in the sense that workers need to gain employment in a core firm at some point in their career in order to flow to a different part of the periphery.

While LFN studies have become increasingly popular, most of this work focuses on descriptive analysis of network topologies.
For example, using employee-employer matched records from the entire economy of Finland, \cite{guerrero_employment_2013} are the first to apply a systematic analysis of LFNs by looking at metrics such as degree distributions, clustering coefficients, associativity, and their relationship to firm properties such as size and profits.
\cite{collet_old_2013} use a similar dataset from Sweden to determine if certain flows can be predicted by the fact that workers were previously employed by the same organisation.
\cite{schmutte_free_2014} applies community detection algorithms to improve the definition of industrial and occupational groups using US survey data.
\cite{guerrero_firmtofirm_2015} deploy configuration models to demonstrate that the matching function of aggregate economic models cannot account for the topology of LFNs.
\cite{park_global_2019} use large-scale LinkedIn data to identify geo-industrial clusters at a global scale. 

Recently, the interest on LFNs has shifted towards a generative modelling point of view.
In particular, the focus has been on understanding the dynamics of labour flows when constrained by an LFN.
For instance, in the same spirit as in econophysics, \cite{guerrero_understanding_2016} develop a random-walk model to study the effect of different network topologies and firm-level parameters in the concentration and dissipation of unemployment after a shock.
\cite{lopez_network_2020} generalise such a model and show that it can predict empirical firm size distributions while enable the inference of firms-specific unemployment.
\cite{delrio-chanona_occupational_2021} applies a similar model to analyse the mobility of workers across occupations. 
\cite{moro_universal_2021} advance upon traditional labour market models by accounting for occupations as sub-markets and updating the job matching function accordingly, using this model to relate the economic resilience of cities to their job connectivity.
\cite{axtell_frictional_2019} go beyond the econophysics approach and provide economic microfoundations to the parameters describing the firms' hiring rates.
They find that, in equilibrium, firm behaviour correlates and amplifies the impact of economic shocks on unemployment. 
\cite{applegate_job_2022} utilise a model based on \cite{axtell_frictional_2019} to explore the relationship between labour mobility, savings, wages, and debt.
To the best of our knowledge, the studies by \cite{axtell_frictional_2019, applegate_job_2022} provide the only models of worker dynamics on LFNs with economic microfoundations.

Unfortunately, both the data analysis and the existing models in the LFN literature rely on one crucial assumption: \textit{that the network is fixed or exogenous}.
This is a reasonable assumption in the study of short-run dynamics,\footnote{\cite{lopez_network_2020} show that, in the short term, there exists a persistent backbone network behind labour flow patterns.} or when one can discard structural transformations.
In other situations such as aggressive technological changes or the shock of a global pandemic, the job and salary landscape may transform in ways that historical labour flows are not a reliable source of information to explain future labour dynamics.
Thus, it is necessary to develop new modelling frameworks that generate observed LFNs endogenously, and go beyond pure stochastic processes by providing economic microfoundations.
This is a challenging task because, in coupling agent-level economic behaviour, one risks losing parsimony and, hence, empirical usability.
In this paper, we develop a model that fits within the proposed framework and overcomes the associated challenges.
In addition, we provide an efficient calibration algorithm, fit the model to comprehensive microdata of UK labour mobility, and systematically analyse the sensitivity of the network topology to restructures in the job and wage distributions across industries, regions, and occupations.
Our approach achieves a balance between the parsimony of econophysics models and the insights into the causal mechanisms provided by agent-level labour market models in order to achieve economic meaningfulness and empirical reliability.\footnote{In the early work by \cite{guerrero_employment_2013}, the authors use a model of firm dynamics developed by \cite{axtell_emergence_1999,axtell_effects_2001,axtell_chapter_2018} to test if it can generate empirical LFNs.
This could be considered the first generative model of LFNs.
They find that Axtell's model can reproduce empirical regularities of LNFs but, due to its high parsimony, it is difficult to reproduce more precise data such as inter-industry flows.
Thus, our aim with in this paper is to move beyond the reproduction of stylised facts, and come closer to the generation of specific LFNs.}

\section*{Results}

\begin{figure}[h!]
	\centering
	\includegraphics[width=1\textwidth]{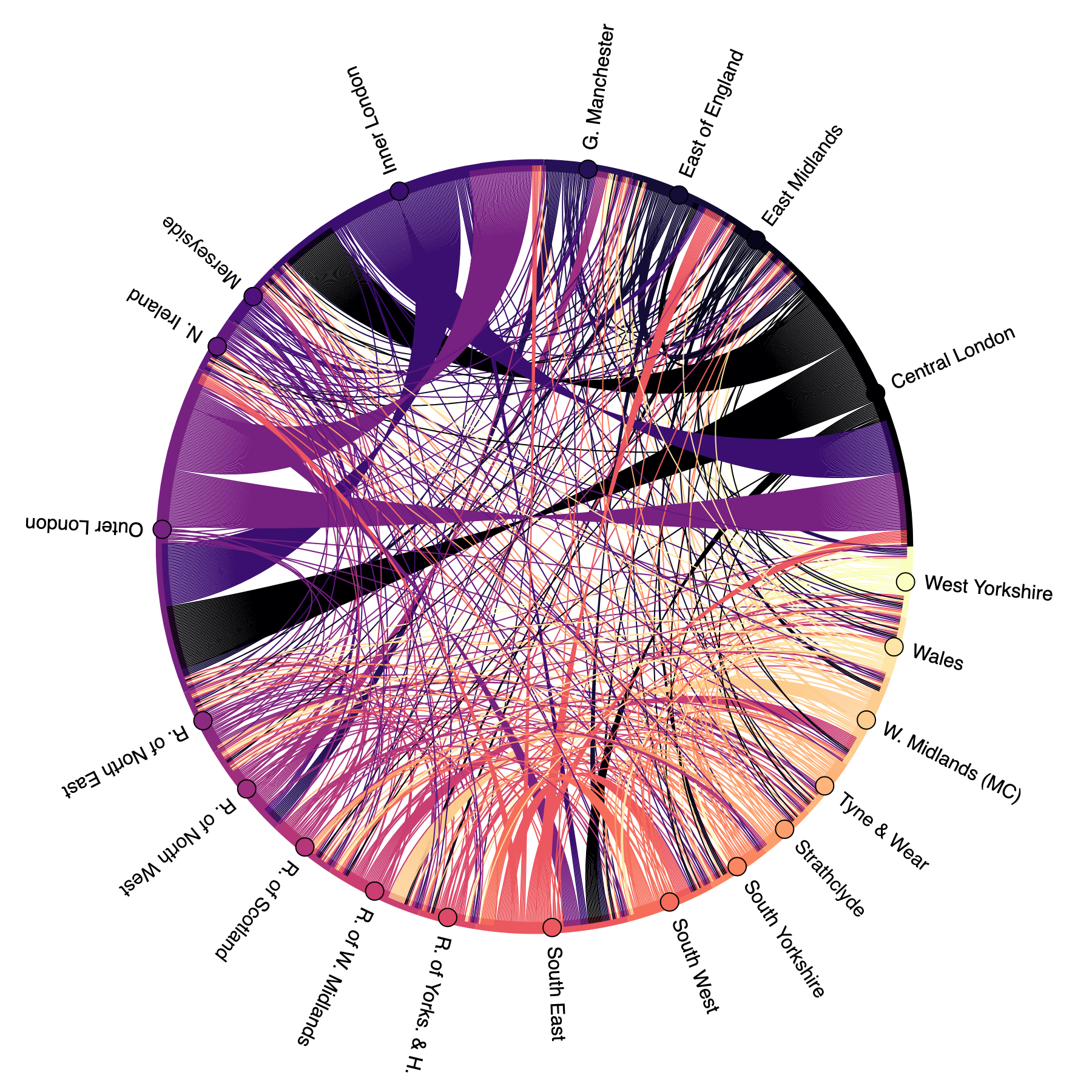}
	\caption{\textbf{Simulated labour flows between geographical regions within the UK.} The plot shows inter-region flows only, as intra-region flows tend to be substantially higher due to the localised nature of job search. The colour of a flow corresponds to the region from which that flow originated. Labels and descriptions for regions are provided in \autoref{tab:region codes}. }
	\label{figure:lfnchordfig_reg}
\end{figure}

\begin{figure}[h!]
	\centering
	\includegraphics[width=0.625\textwidth]{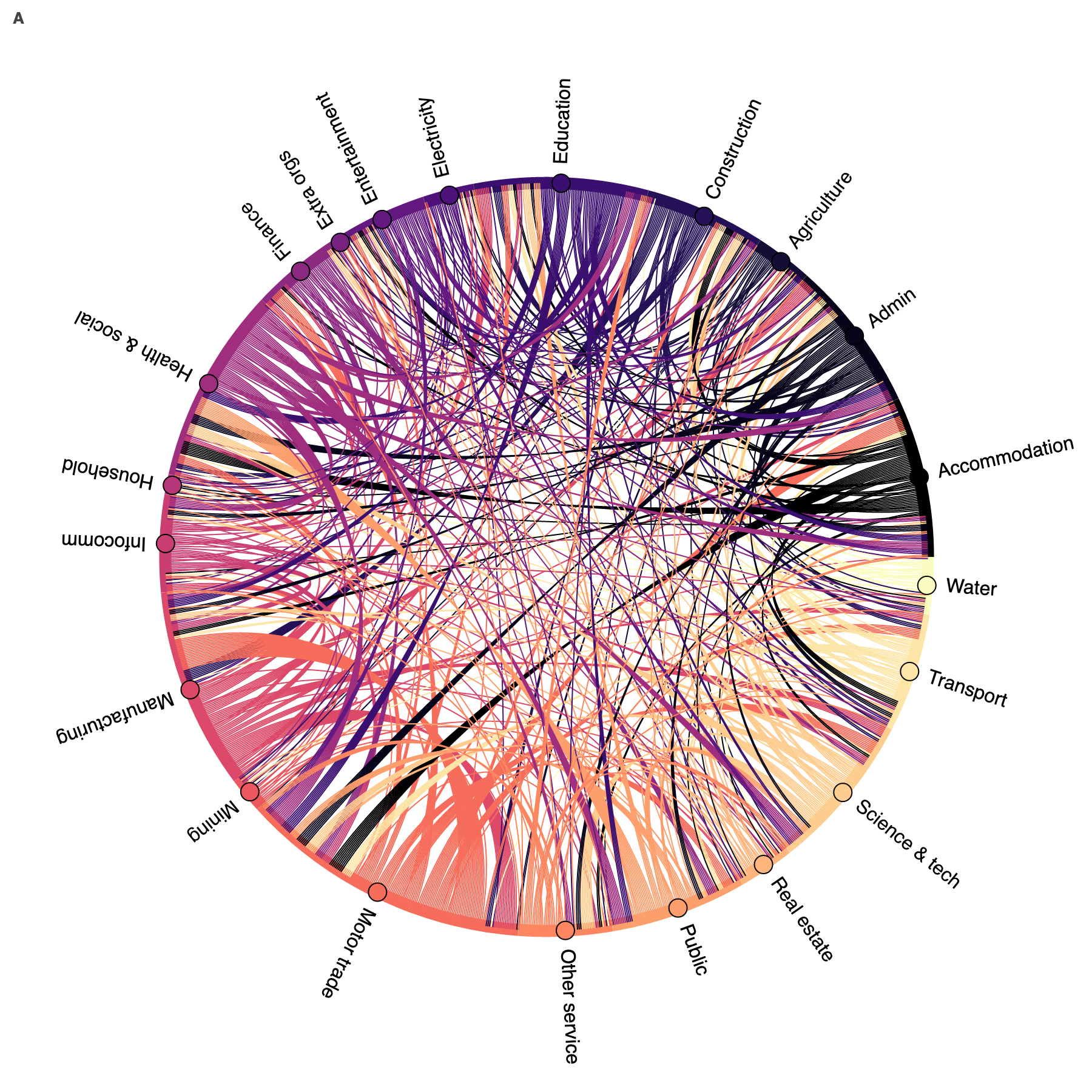}
    \includegraphics[width=0.625\textwidth]{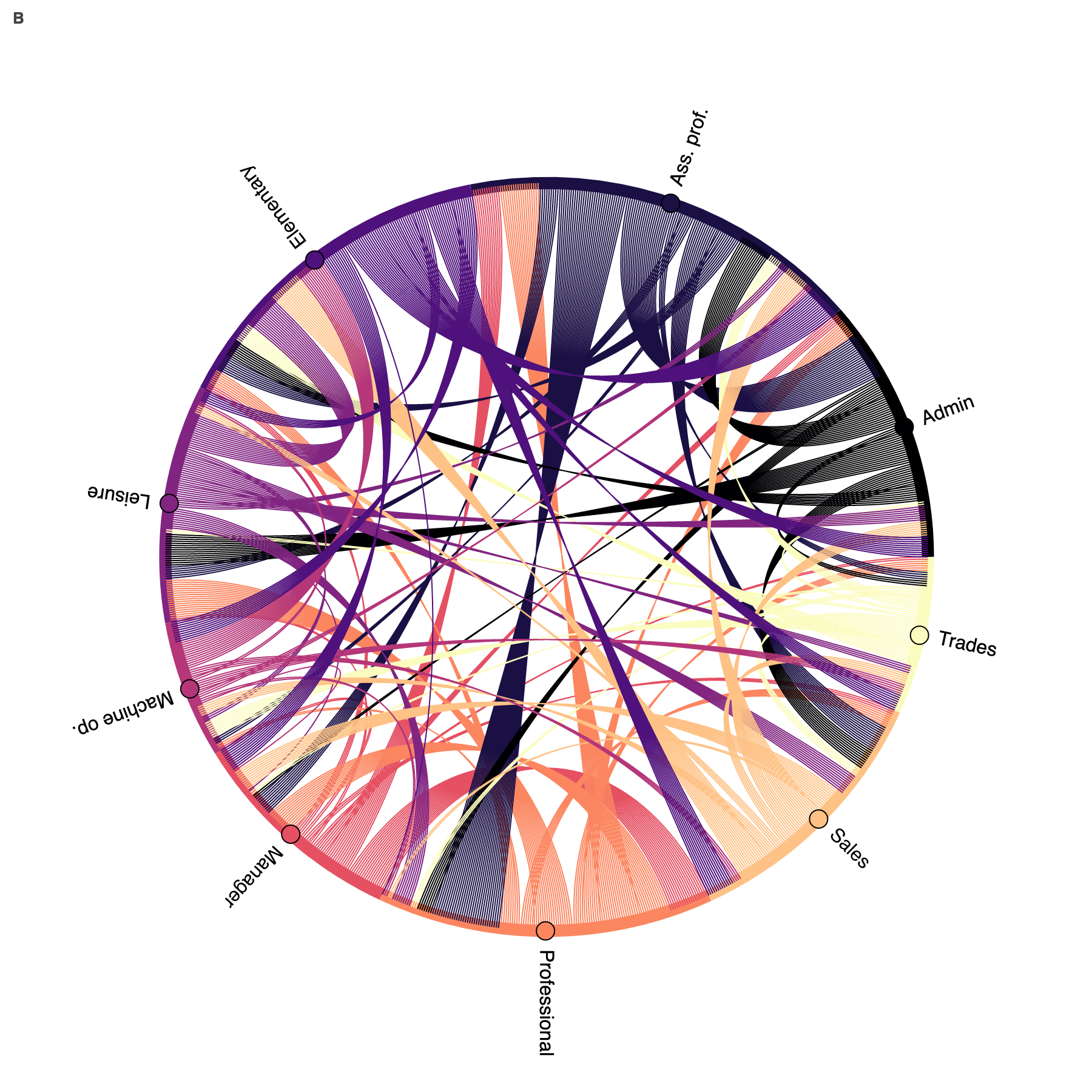}
	\caption{\textbf{Simulated labour flows within the UK.} Subplots indicate inter-group labour flows between a) industries and b) occupations. Only inter-group flows are displayed, as intra-group flows tend to be substantially higher due to the localised nature of job search. The colour of a flow corresponds to the group where that flow originated. Labels and descriptions for industries and occupations are provided in \autoref{tab:industry codes}, and \autoref{tab:occupation codes}. }
	\label{figure:lfnchordfig_suppl}
\end{figure}

Our analysis focuses on UK labour flows between 2012 and 2020.
Due to the coarseness of the data,\footnote{Even if the UK microdata is comprehensive in terms of the number of tracked individuals, it is not designed to capture labour flows in a representative way.
Thus, when defining LFN nodes at  very disaggregate level, the data does not capture a substantial amount of flows.
This is a common problem across all labour force surveys around the world, as they are designed to track individuals, not flows.
Currently, only Nordic-style datasets, which aim at capturing the universe of formal workers through social security records can properly identify the fine-grained structure of LFNs.
Thus, an important task for researchers around the world should be to develop new statistical products designed for capturing LFN structures.} we present our results in terms of relatively aggregate LFNs: one for industries, another for geographical regions, and one more for occupational groups (all part of the same system, not independent of each other).
First, we show how the model is able to reproduce these three LFNs (simulated LFNs are shown in \autoref{figure:lfnchordfig_reg} and \autoref{figure:lfnchordfig_suppl}).
Then, we present results on counterfactual analyses where two types of shocks are introduced: one on the distribution of jobs and another on wages.

\subsection*{Calibration}

The full results from our parameter calibration (refer to Materials and Methods: Calibration) are shown in \autoref{tab:calibration}.
This calibration method estimates all parameters simultaneously and is robust when subjected to changes in scale and to the number of Monte Carlo simulations; it produces consistent results across multiple runs of the calibration method (\autoref{figure:calibration_sensitivity}).\footnote{Additionally, we verify via parameter recovery that the calibration algorithm is able to obtain values that are close to the true ones specified in data-generating model runs  (\autoref{figure:calibration_nus}).}
Here, we summarise the goodness of fit obtained from our calibration procedure across the three types of LFNs using two alternative measures: the Pearson correlation between the empirical and the simulated LFN, and the Frobenius norm.

\begin{table}[h!]
\begin{center}
\begin{tabular}{| p{2.25cm} | p{5.75cm} | p{3cm} |} 
 \hline
 Network & Pearson correlation coefficient & Frobenius norm \\ [0.5ex] 
 \hline\hline
Region &	0.98 (0.32) & 0.05 (1.85)  \\
Industry &	0.96 (0.39) & 0.06 (0.19)  \\
Occupation & 0.93 (0.42) & 0.09 (0.89)  \\
Total &	0.96 (0.38) & 0.07 (0.98)  \\
 \hline
 \hline
\end{tabular}
\end{center}
\caption{\textbf{Correspondence between observed and simulated LFNs.} 
Values presented first in each cell correspond to calculations comparing observed LFNs to those generated from Monte Carlo simulations, while values subsequently presented in brackets correspond to calculations performed to compare observed LFNs with their associated similarity matrix (see Materials and Methods: Calibration - Similarity metrics).
Smaller values for the Frobenius norm indicate better agreement between the two matrices considered.
The row labelled ``Total'' corresponds to a weighted average value taken across the LFNs for region, industry, and occupation.}
\label{tab:calibration}
\end{table}

Across all LFNs, values for the Pearson correlation coefficient and the Frobenius norm (displayed in \autoref{tab:calibration}) indicate a high level of agreement between the observed and simulated job-to-job flows. 
We additionally calculate the same metrics to compare observed job-to-job flows and observed similarity matrices (refer to Materials and Methods: Calibration - Similarity metrics).
In summary, similarity matrices reflect more fundamental constraints that, arguably, would not change so easily with a shock, for example, the essential (most basic) skills required to perform the job.
If there was a high agreement between these similarity matrices and the empirical LFNs, one could say that labour flows are fully explained by these data, and that our model would have little to contribute.
We show the Pearson coefficients and the Frobenius norm in brackets in \autoref{tab:calibration}, and confirm that the simulated flows (i.e. the model output) provide a better fit to the observed flows than the similarity matrices do.
From this, we conclude that the model provides useful information to explain the structure of these LFNs.

\subsection*{Shocks}

We introduce shocks to our simulations to gain an understanding of how changes to the underlying job and wage distributions (such as could result from rapid technological advancement, or a global financial crisis) might alter the job-switching decisions made by individuals, and thus the structure of the LFN.
We consider two types of shocks: one where we alter the occupation and region associated associated with all jobs within a given industry (or set of industries), and another where we alter the wages of all positions within an industry (or set of industries) by shifting the mean wage associated with a group of positions up/down by two standard deviations.
As all positions within a shocked industry are subject to the perturbation, the proportion of total positions shocked depends on how many positions are associated with the perturbed industry (or industries).
We measure the impact of a shock using the weighted Jaccard distance, which indicates the dissimilarity between the flow densities (i.e. the proportion of job-to-job transitions occurring along a given edge of the LFN) in the shocked and unshocked LFNs.
The weighted form is used instead of the unweighted to capture information regarding what fraction of all job-to-job transitions occur along any given edge (i.e. the weight of that edge), not only along which edges job-to-job transitions occur.
In other words, the volume of flow matters, not only the path.

\subsubsection*{Shock size}

\begin{figure}[h!]
	\centering
	\includegraphics[width=1\textwidth]{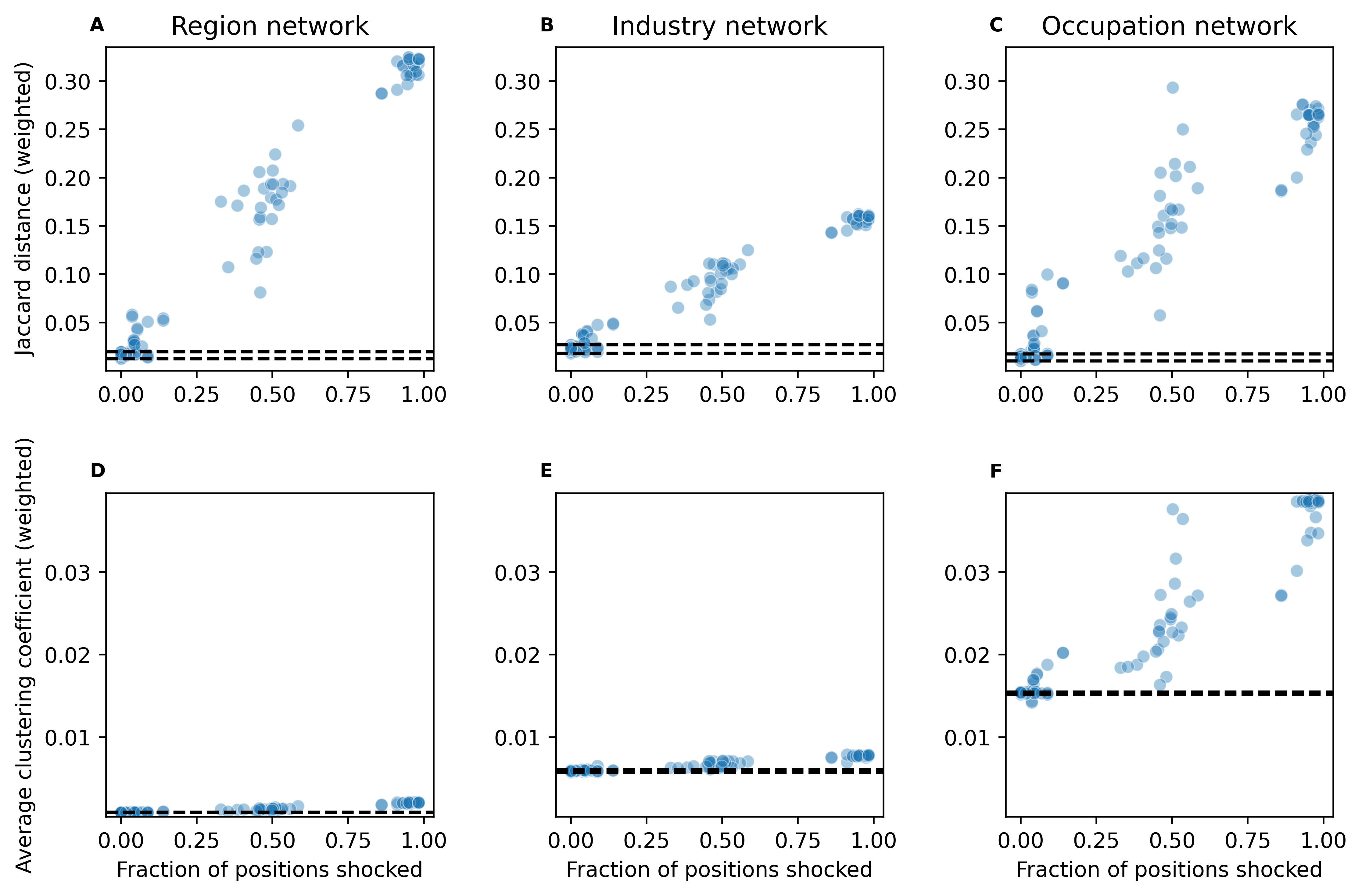}
	\caption{\textbf{Relationship between the size of a shock and its impact on labour flows.}
	Subplots indicate weighted (a-c) Jaccard distance and (d-f) average clustering coefficient values for the region, industry, and occupation LFNs.
	Each point corresponds to the average value taken across a suite of Monte Carlo simulations that all include a shock to the same set of industries. The fraction of positions shocked is calculated as the fraction of all positions that are contained within the set of industries shocked, based on the underlying job distribution.
	Dashed lines indicate the range of values obtained from simulations where no shock has occurred. This variation between simulations in the absence of shocks is a result of the stochastic nature of the model.}
	\label{figure:shocksize}
\end{figure}

When a shock impacts the job distribution, the pattern of flows that emerges as individuals switch jobs differs substantially from the pattern generated from a simulation where no shock has occurred.
This is evident from the value of the weighted Jaccard distance, which indicates the dissimilarity between the flow densities (i.e. the proportion of job-to-job transitions occurring along a given edge) in the shocked and unshocked LFNs (\autoref{figure:shocksize}). 
However even shocks that impact a single industry (and thus only a small proportion of jobs) will, in some cases, cause the Jaccard distance value to rise above the range of values that we would expect to observe when comparing two unshocked networks (\autoref{figure:singleshockmagnitude}). 
We treat this additional distance (i.e. difference between the weighted edge sets) as a result of the shock restructuring the network. 
We also note that shocks increase the weighted average clustering coefficient of the LFNs; i.e. shocks increase the locality of job-to-job transitions, particularly in the case of the occupation LFN.

\subsubsection*{Shock location}

\begin{figure}[h!]
	\centering
	\includegraphics[width=1\textwidth]{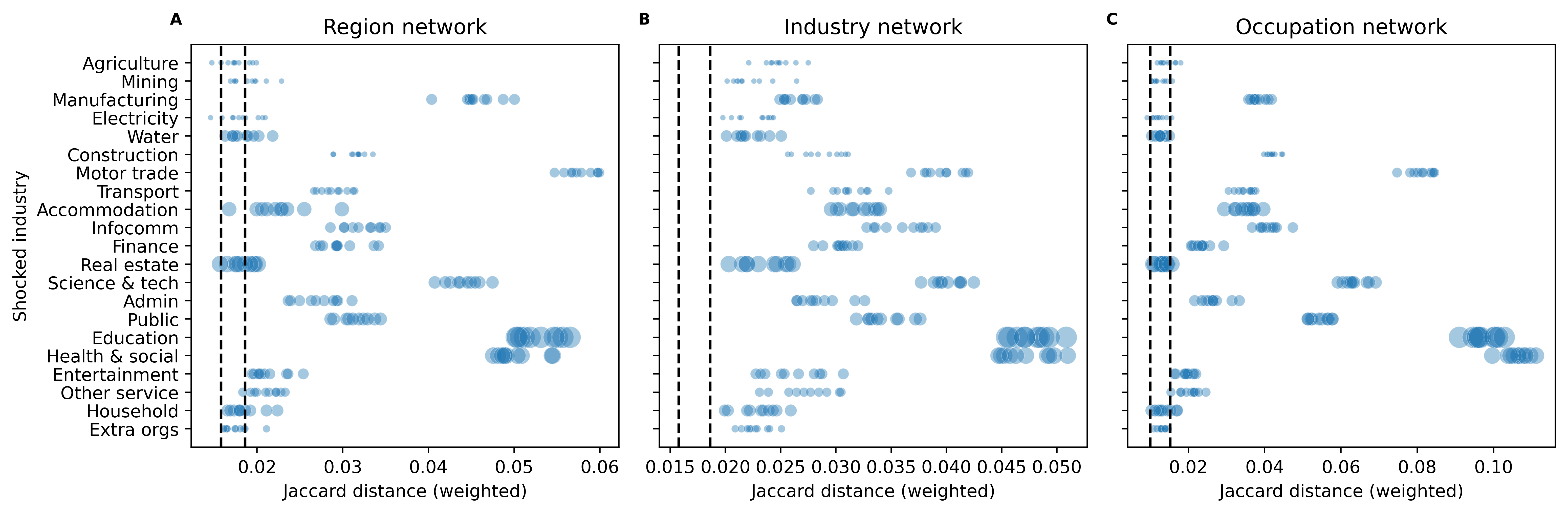}
	\caption{\textbf{Changes to LFN structure resulting from a single-industry shock.}
	Subplots indicate weighted Jaccard distance values for a) region b) industry and c) occupation LFNs obtained by simulating shocks on an industry-by-industry basis.
	Each point corresponds to the average value taken across a suite of Monte Carlo simulations using the same shocked industry, with the size of the point indicating the number of positions contained within the industry.
	Dashed lines indicate the range of values obtained from simulations where no shock has occurred. This variation between simulations in the absence of shocks is a result of the stochastic nature of the model.}
	\label{figure:singleshockmagnitude}
\end{figure}

The extent of shock impacts depends not only on the proportion of jobs that have been shocked, but on which industries have been shocked. 
Industry-by-industry shocks (\autoref{figure:singleshockmagnitude}) confirm this result; shocking motor trade--a relatively small industry --will result in larger Jaccard distance values across all LFNs than a shock to the comparatively large real estate industry. 
Returning to multi-industry shocks, we observe similar results for the occupation LFN.
Shocks impacting roughly 40-50\% of positions and those impacting 80\% or more lead to similar values for the weighted Jaccard distance (roughly spanning 0.15-0.30) and the average weighted clustering coefficient (0.025-0.040) (\autoref{figure:shocksize}C,F). 
However, in general, there is a positive relationship between the fraction of positions impacted and the values of the weighted Jaccard distance and average clustering coefficient.

\subsubsection*{Distribution of shock impacts}

\begin{figure}[h!]
	\centering
	\includegraphics[width=1\textwidth]{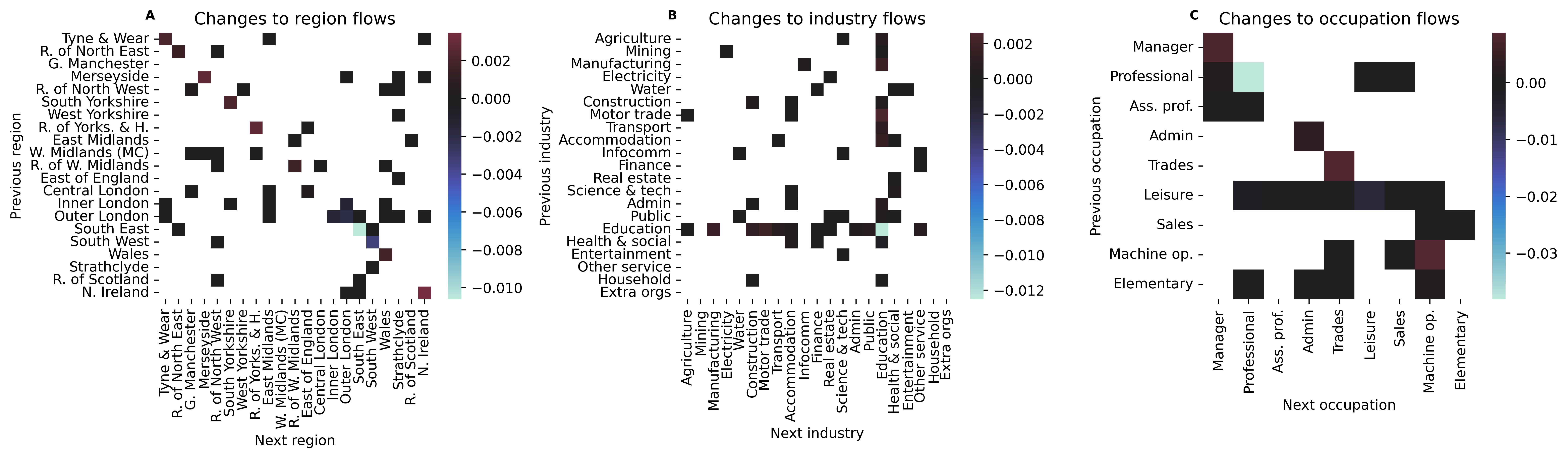}
	\includegraphics[width=1\textwidth]{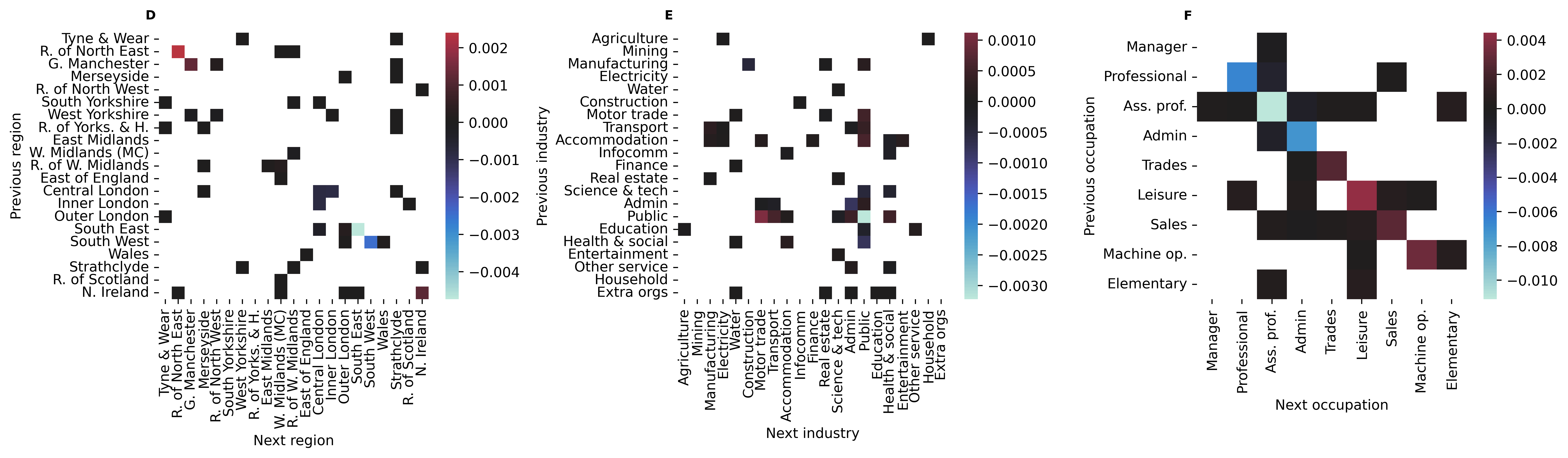}
	\caption{\textbf{Changes in labour flow densities resulting from a shock.}
	Subplots indicate changes to labour flows between/within regions, industries, and occupations for a shock to (a-c) education and (d-f) the public sector.
	Values are calculated by obtaining the mean density across a set of shocked LFNs and a set of unshocked ones both generated by running suites of Monte Carlo simulations, and taking the difference between these two quantities (change = shocked - baseline).
	Only changes to flow densities are larger in magnitude than we would expect to result from model stochasticity - and thus are assumed to be a consequence of the shock - are shown.}
	\label{figure:singleshockflows}
\end{figure}

The way that alterations to job-to-job transition densities (resulting from shocks) are distributed across the LFNs also depends on the specific industry that is being shocked (\autoref{figure:singleshockflows}). 
For some industries (e.g. education) the change in flow density within that industry--in the case of education a substantial decrease in flow density within the industry--in response to a shock is not accompanied by a substantial increase in flow density elsewhere (\autoref{figure:singleshockflows}a-c).
This suggests that numerous industries are ``sinks'' that absorb the workers who, prior to the shock, would have remained within the education industry when switching jobs.
In contrast, when other industries (e.g. public, encompassing public administration, defence, and compulsory social security) are shocked, certain industries (e.g. motor trade) act as the primary ``sinks'' for those transitioning from jobs in the public industry (\autoref{figure:singleshockflows}d-f).
This pattern of multiple sources and sinks is evident across all three LFNs, though for the region and occupation LFNs we primarily see changes to movements within regions (respectively occupations), not between them.
In addition to these differences in how shocks are distributed depending on the industry being shocked, we also observe differences in the distribution of shock impacts across suites of simulations where a given industry has been shocked.

\subsubsection*{Other shocks}

\begin{figure}[h!]
	\centering
	\includegraphics[width=1\textwidth]{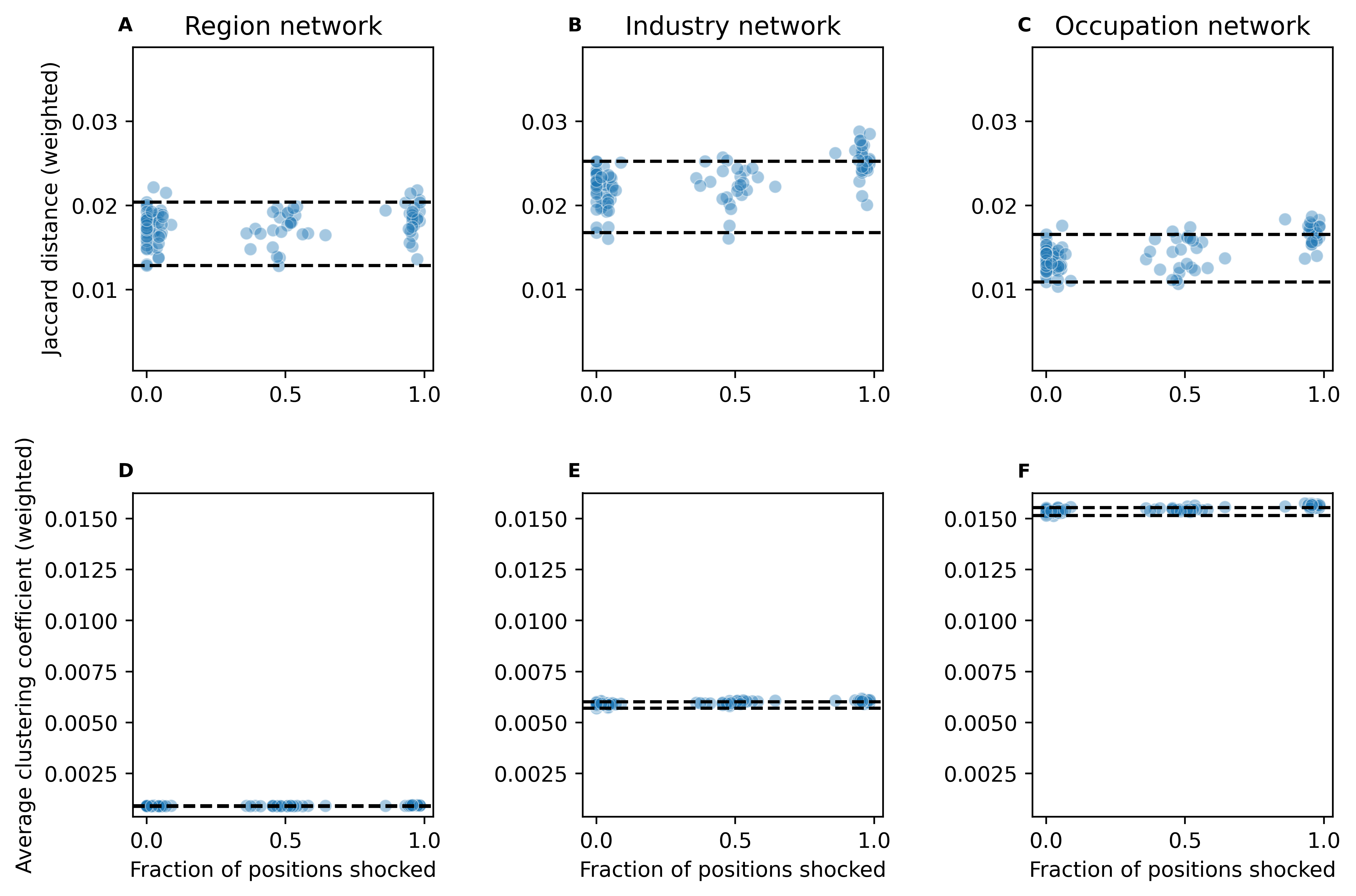}
	\caption{\textbf{Relationship between the size of a shock that decreases average wages and the impact of that shock on labour flows.} Subplots indicate weighted (a-c) Jaccard distance and (d-f) average clustering coefficient values for the region, industry, and occupation LFNs. Each point corresponds to the average value taken across a suite of Monte Carlo simulations using the same set of shocked industries. Dashed lines indicate the range of values obtained from simulations where no shock has occurred. This variation between simulations in the absence of shocks is a result of the stochastic nature of the model.}
	\label{figure:shocksize_wagedecr}
\end{figure}

\begin{figure}[h!]
	\centering
	\includegraphics[width=1\textwidth]{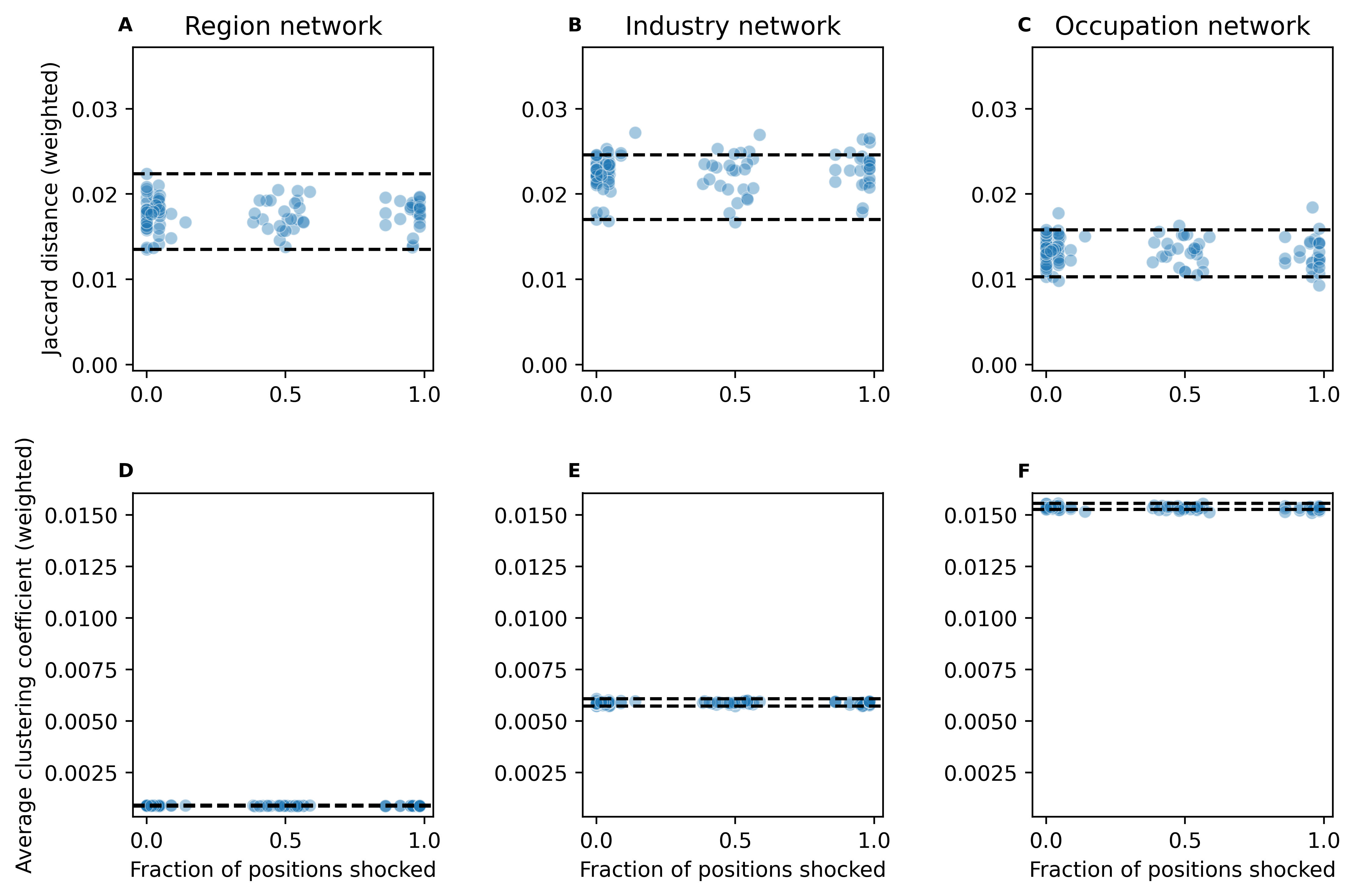}
	\caption{\textbf{Relationship between the size of a shock that increases average wages and the impact of that shock on labour flows.} Subplots indicate weighted (a-c) Jaccard distance and (d-f) average clustering coefficient values for the region, industry, and occupation LFNs. Each point corresponds to the average value taken across a suite of Monte Carlo simulations using the same set of shocked industries. Dashed lines indicate the range of values obtained from simulations where no shock has occurred. This variation between simulations in the absence of shocks is a result of the stochastic nature of the model.}
	\label{figure:shocksize_wageincr}
\end{figure}

We also consider shocks impacting the wages associated with positions in a given set of industries (see \autoref{figure:shocksize_wagedecr}, \autoref{figure:shocksize_wageincr}). 
The impact of these shocks on LFN structure is negligible in comparison to the effect of shocks impacting position characteristics.
In general, the weighted Jaccard distance and average clustering coefficient values for shocked networks do not vary substantially from those values obtained from unshocked networks. 
This is not unexpected; even if an increase (respectively decrease) to the wage associated with a position means that an agent currently receiving a high wage would now apply to (respectively no longer apply to) that position, there will still likely be some lower paid agent on the same node as this high earning agent who will be willing to apply.
As such, we are likely to observe a very similar set of flow densities (but not necessarily micro-level trajectories).
This is mainly due to the broad categories used for our region/industry/occupation characteristics, which may mean that individuals with dissimilar wages are grouped together on the same node.
However, this highlights an aspect of the model not leveraged within this study - the ability to track micro-level trajectories.
If we were to track these paths, it would be possible to gain further insight into the impact of wage shocks by observing which agents (e.g. in terms of their wages) alter their trajectories in response to a shock.
While this falls outside the scope of our current analysis, this avenue of exploration will be pursued in future work.

Our results demonstrate that the UK's LFNs can be replicated with a high level of accuracy using our individual-level model.
This is the first instance in which a model has been used to emerge LFN structure from agent-level behaviour.
Furthermore, we show how changes to the nature of positions (their occupation and region, or the associated wage) impact LFN topology.
These experiments, where LFN topology evolves in response to a shock, move beyond previous methodologies where \textit{ad hoc} assumptions needed to be made about how LFN structure would be altered by the introduction of a shock.

\section*{Discussion}

This paper advances upon previous studies of LFNs by introducing an agent-level model that emerges LFNs from micro-level behaviour.
The model, once calibrated, is used to explore changes to LFN structure resulting from the restructuring of underlying job and wage distributions, something that demands attention in a time of a substantial restructuring in many labour markets.
As alterations to the wage distribution have little impact on LFN structure (though they may effect worker income and other micro-level outcomes), we focus on positional restructuring. 
These changes alter job-to-job flow densities and increase the average clustering within the LFN. 
The magnitude of the resulting impacts increases with the proportion of positions affected. 
However, the magnitude and distribution of impacts are not solely dependent on the number of positions affected, with considerable variation in outcomes when changes are applied to industries containing similar numbers of positions. 
For example, shocking the Health \& social sector has a much larger impact on job-to-job flow densities across the region, industry, and occupation LFNs than shocking the Real estate sector, despite both containing a similar number of positions.
These results highlight the need for models that endogenously emerge network structures.

This modelling framework opens up numerous possibilities for exploring labour market dynamics. 
By removing the need to assume a static network structure, it can be used to provide insights into plausible trajectories along which the labour market might evolve under different scenarios (e.g. a period of rapid technological change, the establishment of new international trade agreements, etc.)
Additionally, as the model is built on agent-level behaviour and leverages large-scale microdata, it provides highly granular insights into career trajectories to inform evidence-based policymaking. 
Due to the flexible nature of the modelling framework, it has broad applicability beyond the UK context we have focused on.
Furthermore, because of its micro-level economic specification, it allows to analyse shocks and interventions different from the ones discussed in this piece, e.g. income taxes, skill-up policies, social transfers, unemployment and matching programmes, etc.
The model--which is already quite parsimonious--can be adapted to accommodate what data is available.
These data could come from the labour force surveys administered in other countries.
In particular, there is great flexibility in terms of the level of aggregation of the characteristic variables (geography, industry, occupation).
As such, a broad spectrum of data can be admitted into the model, from the highly disaggregate data found in the labour force surveys of Nordic countries, to the more aggregate data collected in developing nations.
Additionally, model structure and agent behaviour can be altered should the user wish to explore some aspect of labour markets (e.g. job precarity, or the impact of skills on career trajectories) that the model does not currently account for.

\section*{Materials and Methods}

\subsection*{Data}

The main data source is the UK's Labour Force Survey (LFS), consisting of longitudinal information about British households and individuals \cite{officefornationalstatistics_labour_2022a}. 
We utilise data from 2012-2020, collected on a quarterly basis, with individuals being sampled for 5 consecutive waves, and a fifth of the sample being replaced every wave. 
Currently, each quarterly dataset (covering 5 waves) contains information on approximately 37,000 households (90,000 individuals) from Great Britain and Northern Ireland \cite{officefornationalstatistics_labour_2022}.

The LFS identifies the region, industry, and occupation of the respondent's job at the time of the interview. 
It can thus be used to track changes in these variables over time such that regional, industrial, and occupational mobility can be measured and modelled. 
These data allow us to identify job-to-job flows, where employed individuals move to a new position. 
Several individual characteristics are also present, for example, sociodemographic ones (age) and employment-related variables (net weekly earnings, hours worked, etc.)

The flows identified in the LFS data characterise the complex mobility patterns that emerge in the labour market through time. 
LFNs have become a standard way to encode such information in an intuitive and analytically tractable way \cite{guerrero_employment_2013, guerrero_firmtofirm_2015, axtell_frictional_2019}.
We construct three LFNs by accumulating observed labour flows from the LFS data, and grouping these flows by geographical regions (based on the GORWKR variable from LFS, hereafter referred to as regions), industries (UK Standard Industrial Classification sections), or occupations (UK Standard Occupation Classification major groups). 
Therefore, these LFNs are directed weighted graphs where each node represents a region (or industry, or occupation) and an edge between two nodes represents a flow of labour from one region (or industry, or occupation) to another. 
If a node were to represent a combination of these three characteristics, the data would yield LFNs that are too sparse, making them unsuitable for model calibration. 
Nevertheless, with more comprehensive administrative data such as employee-employer matched records--like the ones held by social security agencies and treasuries--this sparsity problem could be overcome.




\subsection*{Model}

\begin{figure}[h!]
	\centering
	\includegraphics[width=1\textwidth]{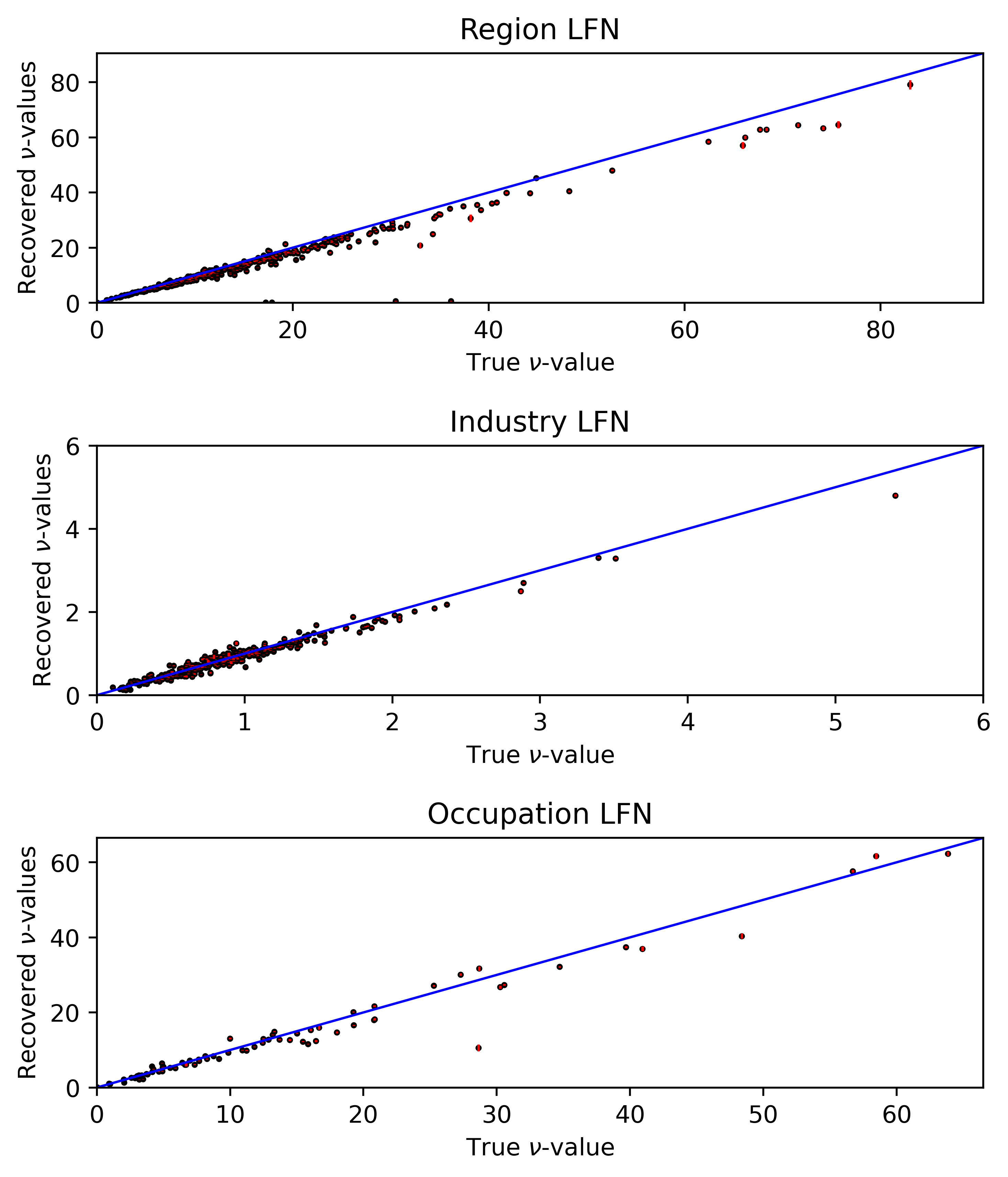}
	\caption{\textbf{Results of parameter recovery exercise.} For each $\nu$-value, we make 100 attempts to recover the true value by running the calibration algorithm with the objective of fitting to the LFNs generated by a simulation using the true $\nu$-values. For each set of 100 recovered $\nu$-values the mean is indicated with a black point, and the 99\% confidence interval is indicated with a red vertical line. The $y=x$ line is indicated in blue. For each run of the calibration algorithm, a suite of Monte Carlo simulations was performed with $N=3500$.}
	\label{figure:calibration_nus}
\end{figure}

Our model aims to achieve a balance between the parsimony of stochastic processes \textit{à la} econophysics and the microfoundations of economic models.
This balance is important because the econophysics approach allows highly disaggregated analysis of systems with heterogeneous and rationally-bounded agents, while agent-level microfoundations set socially meaningful causal mechanisms through which these agents respond to the economic environment.
In particular, we construct a model where agents optimise leisure and consumption in a myopic fashion.
Myopia, as opposed to solving infinite horizon inter-temporal optimisation problems, produces disequilibrium at the micro-level as agents continuously find themselves in situations with incentives for unilateral deviations from the status quo (triggering a stream of labour flows).\footnote{In other words, there is no price vector that coordinates their actions in a centralised fashion and generates an equilibrium, a well accepted principle in the complexity economics literature \cite{tesfatsion_agentbased_2003, axtell_what_2007, kirman_complex_2010, arthur_complexity_2009}.}
Despite micro-level disequilibrium, the aggregate dynamics reach steady-state behaviour, and this facilitates model calibration.
We develop a multi-output stochastic gradient descent algorithm for this purpose, and show (\autoref{figure:calibration_sensitivity}) that the calibration method is robust to variation in the number of agents simulated (except in the trivial case of a very low number of agents), and to the number of Monte Carlo simulations performed.
We also show that calibration results are able to recover the true parameters of the model when used on simulated data (\autoref{figure:calibration_nus}).

Our model consists of $N$ heterogeneous agents representing individuals within the labour market.
Each agent has an age, consumption preferences, and budgetary constraints, which they consider when deciding whether to apply or not to vacant job positions.
These characteristics are drawn from LFS microdata. 
With each simulation period, agents age and may die with a probability $1- \omega$, where $\omega$ is a function of their current age.
Dead agents are replaced by new ones, aged 18, with all other characteristics randomly drawn from their empirical distributions.
Positions are individually modelled as objects with characteristics such as wage, industry, region, and occupation.
They are exogenous and are created and destroyed at a constant rate (such that filled positions that are destroyed generate unemployed agents).
The agents search for and apply to new positions (both while being unemployed and employed). 
The probability of applying to a position depends on certain similarity properties (that we explain ahead) between the new position and the current (or previous in the case of unemployment) one, as well as the wage differential.
The likelihood of being hired, on the other hand, depends on the affinity of competing applicants.
The number of agents and positions remains constant and we focus on studying the impact of two sources of structural change in the LFN topology: a redistribution of job positions and a change to the mean wage, across industries, regions, and occupations. 
This focus is informed by the large body of evidence suggesting that shocks to the economy (e.g. advances in artificial intelligence and automation, the COVID-19 pandemic \cite{_four_, frank_understanding_2019, _amazonian_, _impact_}) alter the types of jobs that are available, where they are located, and the wages associated with them.
Next, we provide the full details of each model component.

\subsubsection*{Job creation and destruction}

Suppose there are $P$ job positions in the economy.
Each position has the following attributes: a wage, a region, an industry, an occupation, and a state (vacant or filled).
If vacant, the position is not associated with any agent.
If filled, the position becomes associated with the agent that performs the job.
Every step, each position (regardless of whether it is filled or vacant) can be destroyed with a probability $\lambda$.
Parameter $\lambda$ is both the job destruction and the job creation rate so, in the same step, $\lambda P$ positions are destroyed and subsequently replaced with new ones.
The job creation/destruction rate and wages are assumed to be exogenous.\footnote{While the reader may wish to make $\lambda$ or the wages endogenous variables, this would require additional assumptions (e.g. about firm behaviour) and data.
Such adjustments complicate the model in ways that we consider unnecessary for the analysis conducted in this paper.}

\subsubsection*{Agents}

Agents can be in one of two states: employed or unemployed.
They are always associated to the region, industry, and occupation of their current (if employed) or last (if unemployed) position (following \cite{guerrero_understanding_2016, axtell_frictional_2019}).
Agents whose positions are destroyed become unemployed.

The agent behavioural component follows the classic leisure-consumption economics textbook model.
In period $t$, agent $i$ benefits from $l$ time units of leisure (where $l \in [0,1]$, and $1-l$ denotes the time units devoted to work) and $c$ units of consumption through a utility function

\begin{equation}
 U_{i,t} \left(c_{i,t}, l_{i,t}\right)=c_{i,t}^{\alpha_i} l_{i,t}^{1-\alpha_i}
 \label{eqn:utilityfcn}
\end{equation}
where $\alpha_i$ is a consumption preference parameter.
The intuition behind this standard model is that agents receive utility from the consumption that they can afford through their labour income, and from the leisure time they can spend by not working.
The trade-off between working to consume and not working to gain more leisure time is partly determined by the preferences of the agent, which are modelled through parameter $\alpha_i$.
Hence, a higher $\alpha_i$ means that the agent would prefer to work more, as this would enable higher consumption levels.
As is customary, we assume that $0 < \alpha_i < 1$.

This leisure consumption model has a more general form that considers inter-temporal choices.
For simplicity, we assume that these choices do not involve savings, so all the income earned in period $t$ is spent.
This allows us to write the inter-temporal utility function as the series 

\begin{equation}
    U_i = \sum_{t=0}^\infty \gamma^t  U_{i,t} (c_{i,t},l_{i,t}),
    \label{eqn:futureutility}
\end{equation}
where $\gamma$ is a discounting factor.
Since consumption choices are time independent, we can rewrite the previous series as

\begin{equation}
   U_i = \frac{\gamma^{L_i}}{1 - \gamma} c_i^{\alpha_i} l_i^{1 - \alpha_i},
    \label{eqn:futureutility_compact}
\end{equation}
where $L_i$ is the age of the agent, so utility is age-dependent as individuals tend to discount future utility differently as they age.

\autoref{eqn:futureutility_compact} may exaggerate the cognitive capabilities of agents since such calculations in an infinite horizons may seem unrealistic.
However, it offers the benefit of replacing computationally intensive (simulation-wise) numerical algorithms with a single calculation, which is important for the computational efficiency of the model.

Next, let us introduce the budget that constrains the consumption choices of the agent.
Let $w_i$ denote the wage (per time unit) received by an employed agent (remember that this is a feature of the position).
Every period, all this income is used for consumption purposes, so we say that the identity $c_i = (1-l_i) w_i$ holds.
Then, provided the utility function and the budget constraint, the agent solves the maximisation problem

\begin{equation}
\begin{split}
\max_{c_i,l_i} U_i = \frac{\gamma^{L_i}}{1 - \gamma} c_i^{\alpha_i} l_i^{1 - \alpha_i}\\
\text{s.t.} \quad c_i = (1-l_i) w_i
\end{split}\label{eqn:utilitymaximisation}
\end{equation}

While we may sacrifice realism in terms of assuming utility-maximising agents, we emphasise it in how these decisions are difficult to perfectly couple them with employment choices.
Agents are rationally bounded in the sense that they do not produce sophisticated expectations on the trajectories of future utility such as considering different employment scenarios (i.e. calculating probability distributions on all the potential job opportunities that may arise and the corresponding employment outcomes). 
Instead, we assume that the leisure-consumption problem is separable from the one of choosing whether to stay in the same job or to take on a new position.
Thus, when solving the utility maximisation problem, agents assume their wage are fixed.
However, in the potential case of being presented with a job offer, they are able to form a counterfactual scenario with the potential new wage and to compare the utility differential.\footnote{This specification also implies that the agent assumes they will remain employed in the future.
This could be adjusted to introduce myopic employment expectations, for example, from information coming from social networks or by considering heterogeneous separation rates (as currently $\lambda$ is homogeneous so accounting for it does not affect our analysis).
For this paper, we omit employment expectations as they are not the focus of this study and they would require additional data to calibrate any associated parameters.}

To finalise the agent component of the model, the solution of the agent's optimisation problem in period $t$ is

\begin{equation}
    U_{i,L_{i,t}}^* = \frac{\gamma^{L_{i,t}}}{1-\gamma} w_{i,t} \alpha_i^{\alpha_i} \left(\frac{1-\alpha_i}{w_{i,t}}\right)^{(1-\alpha_i)}.
    \label{eqn:optimalutility}
\end{equation}

\subsubsection*{Matching, application, and hiring}

The probability of a match between a position and an agent depends on the industrial, regional, and occupational similarity between the vacant position and the agent's current position (or previous position in case of an unemployed worker).
These similarity metrics or scores are weighted through free parameters that we calibrate to match the empirical LFN.
The idea is that these similarities and their weights capture more fundamental elements of the labour market than the idiosyncratic factors that may be reflected in observed flow data.
Thus, should a substantial change in the distribution of jobs or wages take place, this approach would allow for a new network topology to emerge.

Let us consider an agent employed in position $k$.
At any given period $t$, this agent may be matched to another position $g$ with a probability

\begin{equation}
    \xi(k,g) \propto S(k,g),
    \label{eqn:matchprobability}
\end{equation}							
where $S(k,g)$ is the degree of similarity between the two positions.

Once a match takes place, the agent applies to position $g$ only if they expect to gain more utility from working in the new job. 
The agent's choice is determined by a utility comparison.
Let $w_k$ and $w_g$ denote the wages associated to positions $k$ and $g$ respectively.
To make a decision, the agent assesses their future utility by calculating the counterfactual with a new salary and comparing it to the current utility outcome.
This choice is expressed through the condition

\begin{equation}
   U_{i}^*(g) >U_{i}^*(k). 
   \label{eqn:utilityinequality}
\end{equation}
Thus, if the future utility stream under the new job is larger than the one from the current job, then the agent has incentives to switch jobs.
For simplicity, we assume that wage $w_g$ is public information and that, if \autoref{eqn:utilityinequality} is satisfied, the agent will apply to position $g$.

After all job searchers have submitted applications, each vacant position ranks its applicants; first, according to the similarity score $S_{k,g}$ and, then, by the order of submission. 
Following this, a sequential process of making offers takes place. 
Here, the vacant positions (in random order) hire the best-ranked applicant.
Therefore, there may be positions that remain vacant because they are unable to attract suitable candidates (generating frictional unemployment).
For agents who are currently unemployed, the position from their most recent employment spell is used to calculate $S_{k,g}$ for both the matching protocol and the applicant ranking process.

\subsubsection*{Job search intensity}

Consider that, with probability $\mu_{i,t}$, agent $i$ participates in the matching protocol in period $t$.
We define this probability in terms of the agent's current employment status, assuming that the likelihood of actively searching differs based on this status, such that

\begin{equation}
\mu_{i,t} =
    \begin{cases}
    \theta_e \ \text{if agent $i$ is employed in period $t$}, \\
    \theta_{ue} \ \text{if agent $i$ is unemployed in period $t$}.
    \end{cases}
    \label{eqn:jobsearch}
\end{equation}													
This search intensity is also known in the agent-computing literature as the activation rate: the probability that an agent is active to engage in interactions during a given period \cite{axtell_effects_2001}.

\subsection*{Model summary}

Here we provide a brief overview of the model parameters (\autoref{tab:parameters}) and of the dynamic processes occurring during each period within a simulation (\autoref{alg:pseudocode}).
For parameters representing agent attributes, when a new agent is created, a random value is drawn for each attribute from the empirical marginal distributions.
Overall, the model is quite parsimonious since its equations are directly interpretable, the causal channels are explicit, and there are few free parameters.

\begin{table}[h!]
\begin{center}
\begin{tabular}{| p{1.75cm} | p{4.5cm} | p{2cm} | p{2cm} | p{2.5cm}|} 
 \hline
 Parameter & Description & Nature & Source & Diversity \\ [0.5ex] 
 \hline\hline
$N$ &	number of agents &	endogenous & 	\cite{theworldbank_labor_2022} & NA \\
 \hline
$P$ &	number of positions &	endogenous &	\cite{officefornationalstatistics_vacancies_2020} &	NA \\
 \hline
$\omega$ &	survival probability (age-specific) &	exogenous &	\cite{officefornationalstatistics_national_2021} &	heterogeneous \\
 \hline
$\alpha$ &	consumption preference &	exogenous &	\cite{officefornationalstatistics_labour_2022a} &	heterogeneous \\
 \hline
$\gamma$ &	discount rate &	exogenous &	\cite{hmtreasury_green_2022} &	homogeneous \\
 \hline
$w$ &	net wage (annual) &	partially exogenous	 & \cite{officefornationalstatistics_labour_2022a} &	heterogeneous \\
 \hline
$L$ &	agent age	 & partially exogenous &	\cite{officefornationalstatistics_labour_2022a} &	heterogeneous \\
 \hline
$\lambda$ &	job creation/destruction rate &	exogenous &	\cite{officefornationalstatistics_business_2020} &	homogeneous \\
 \hline
$S$ &	similarity metric &	exogenous &
calibration, \cite{officefornationalstatistics_national_2021,u.s.departmentoflabor_net_2022,ukdepartmentforeducation_lmi_2022,openstreetmapfoundation_openstreetmap_2022} &	heterogeneous across nodes \\
 \hline
$\theta_{e(ue)}$ &	activation rate for employed (respectively unemployed) individuals &	exogenous &	\cite{officefornationalstatistics_labour_2022a} &	homogeneous \\
 \hline
$U$ &	utility &	endogenous &	NA &	heterogeneous \\ 
 \hline
\end{tabular}
\end{center}
\caption{\textbf{Model parameters.}
Variables with a partially exogenous nature are those where the initial value is exogenously determined through a random draw from the marginal distribution, but its evolution is endogenously determined by the model.
Unless specified otherwise, heterogeneity is measured with respect to the agents.}
\label{tab:parameters}
\end{table}

\begin{algorithm}
    \SetKwInOut{Input}{Input}
    \SetKwInOut{Output}{Output}

    \Input{$\{N, P, \omega, \alpha, \gamma, w, L, \lambda, S \}$ for every agent $i$}
    \ForEach{period $t$}
      {
        destroy positions and create new unemployed\;
        create new vacant positions\;
         \ForEach{agent $i$}
            {
            increase agent's age\;
            \If{agent $i$ is active}
                {
                sample vacant position $g$\;
                \If{$U_{i}^*(g) >U_{i}^*(k)$}
                    {
                    agent $i$ applies to vacant position $g$\;
                    }
                }
            }
        \ForEach{vacant position $g$}
            { rank applications\;
             hire the best applicant\;
             }
          \ForEach{agent $i$}
            {
            perform survival trial\;
                \If{agent $i$ dies}
                    {
                    remove agent $i$ from the population\;
                    create a new agent
                    }
            }
      }
    \caption{Model pseudocode} \label{alg:pseudocode}
\end{algorithm}

\subsection*{Calibration}

To instantiate the agent population and run the model, it is necessary to assign values to the model parameters.
This is done via a combination of (1) direct imputation from microdata and (2) calibration.

\subsubsection*{Direct imputation}

For the macro-level parameters, we have the number of agents $N$ and of positions $P$. 
In 2019 there were approximately 35,000,000 individuals within the UK labour force \cite{theworldbank_labor_2022}, and the number of job positions in the UK was roughly 36,000,000 with around 800,000 of these standing vacant \cite{officefornationalstatistics_vacancies_2020}. 
Due to the computational costs of calibrating the model at full scale, we specify $N$ and $P$ at a 1:10,000 scale.
This scale is selected, as moving to a 1:1,000 (or finer) scale does not lead to an accompanying increase in the level of accuracy attained during calibration (\autoref{figure:calibration_sensitivity}).
The job creation/destruction rate is set to $\lambda = 0.0463$ by averaging the job creation and destruction rates in the UK from 2011-2019 \cite{officefornationalstatistics_business_2020}. 
Age-stratified survival probabilities are obtained from the UK Office for National Statistic's Life Tables for 2017-2019 \cite{officefornationalstatistics_national_2021}. 
The discount rate $\gamma=0.9662$ is taken from HM Treasury's Green Book \cite{hmtreasury_green_2022}.
The Green Book is a document providing guidance to ministers within the UK government as to how to achieve policy objectives \cite{_final_}. 
 
The initial age $L_{i,0}$ of each agent is taken directly from the LFS microdata \cite{officefornationalstatistics_labour_2022a}. 
Several other parameter values are imputed from this microdata.
The preference parameter $\alpha_i$ is imputed through the identity $\alpha_i=\frac{c_i}{w_{i,t} l_i+c_i}$ from the first-order conditions of the utility maximisation problem presented in \autoref{eqn:utilitymaximisation}. 
Activation rates for employed (respectively unemployed) agents are determined by computing the fraction of employed (respectively unemployed) individuals who are actively searching for a new job. 
Every time a new position is created, a random wage is generated from a normal distribution with mean equal to the empirical mean wage of the industry-region-occupation of the position and with standard deviation equal to the empirical standard deviation of wages associated with the industry-region-occupation of the position.\footnote{For nodes where there are at least 10 wage observations available in the LFS microdata, the assumption of normality has been checked using a Kolmogorov-Smirnov test (at a 5\% significance level).
For nodes with fewer than 10 associated wage observations, we need to assume normality.}

\subsubsection*{Similarity metrics}

The similarity ($S_{k,g}$) between positions $k$ and $g$ is determined by their geographical proximity, industrial affinity, and occupational closeness. 
To measure the geographical proximity between two regions, we employ the inverse of the physical (great-circle) distance between their major urban centres \cite{openstreetmapfoundation_openstreetmap_2022}. 
Within the regional similarity matrix $R$, entry $R_{i,j}$ contains values given by the geographical proximity between regions $i$ and $j$

\begin{equation}
    R_{i,j} = 1 - \frac{d_{i,j} - \min \left(d_{i,j}\right)_{\forall i, j}}{\max \left(d_{i,j}\right)_{\forall i, j} - \min \left(d_{i,j}\right)_{\forall i, j}},
    \label{eqn:regionsimilarity}
\end{equation}
where $d_{i,j}$ denotes the physical distance between $i$ and $j$. 
This specification is consistent with the literature on urban economics and human geography, where gravity models are built on the same principle.

The affinity between two industries is determined by the relative importance of one of the sectors as a supplier of the other. 
To quantify this, we employ the UK's Input-Output Analytical Tables from 2018 \cite{officefornationalstatistics_uk_2022}. 
Within the industrial similarity matrix $I$, the affinity between industry $i$ and $j$ is defined by

\begin{equation}
    I_{i,j} = \frac{x_{i,j}}{\sum_{k=1}^{n_i} x_{i,k}},
    \label{eqn:industrysimilarity}
\end{equation}
where $x_{i,j}$ is the value of the resources produced by industry $j$ that are being used as inputs for industry $i$ and $n_i$ is the total number of industries. 
This metric implies that the affinity scores are not symmetric. 
Thus, the direction of the employment path of an agent influences their future mobility, which is consistent with the directed nature of the LFNs.

Finally, we determine how close two occupations are by considering their skill composition. 
The O*NET \textsuperscript{\textregistered} system quantifies the level of a given skill required to perform a given occupation \cite{u.s.departmentoflabor_net_2022}. 
While O*NET \textsuperscript{\textregistered} is built for the US, the Labour Market Information for All API provides a mapping from the SOC codes to their O*NET \textsuperscript{\textregistered} counterparts \cite{ukdepartmentforeducation_lmi_2022}. 
To construct a closeness score between two occupations we apply the following steps:

\begin{enumerate}
    \item Collect data on the skills associated with each SOC code.
    If the SOC code covers more than one O*NET \textsuperscript{\textregistered} code, we take the average skill level across all O*NET \textsuperscript{\textregistered} codes.
	\item Group the skills associated with these SOC codes by their SOC major group and calculate mean values for each skills' level to generate a skill-vector for each occupation (at the SOC major group level).
	\item Calculate $O_{i,j}$, the closeness between occupations $i$ and $j$, given by the cosine similarity of their skill-vectors as follows:
\begin{equation}
    O_{i,j} = \frac{\sum_{k=1}^{n_s} \pi_k^i \pi_k^j}{\lVert \pi^i \rVert \lVert \pi^j \rVert},
    \label{eqn:occupationsimilarity}
\end{equation}
where $\pi^i$ is the skill-vector of occupation $i$; $n_s$ is the number of different skill categories (e.g. the length of the skill-vectors); and $\lVert \cdot \rVert$ denotes the $L^2$ norm. 
\end{enumerate}

All three matrices are then normalised such that their entries fall within [0,1].
Using these matrices, we construct $S_{k,g}$, the measure of similarity between positions $k$ and $g$ as

\begin{equation}
    S_{k,g} = R_{k,g}^{\nu^R_{k,g}} I_{k,g}^{\nu^I_{k,g}} O_{k,g}^{\nu^O_{k,g}},
    \label{eqn:similarity}
\end{equation}
where $\nu^R, \ \nu^I, \ \nu^O$ are matrices of parameters for weighting the importance of geographical proximity, industrial affinity, and occupational closeness to defining the similarity between two positions. 
These are free parameters and must be calibrated.

\subsubsection*{Calibration algorithm}

Calibration is performed to determine values for the parameters contained in the matrices $\nu^R, \ \nu^I, \ \nu^O$. 
This calibration procedure is performed using only data generated in the steady state of the model. 
First, each simulation is run for a sufficient length of time to reach the steady state (where the network of flows has stabilised).\footnote{It is reasonable to gauge the time needed to reach a steady state by observing the time-series output of the calibrated model and subsequently running simulations for a sufficient length of time.
However, should the reader wish to more rigorously define the steady state, please refer to Supplementary Information: Defining a steady state.} 
Following this, the simulation runs for an additional period of time at its steady state (enough for the network of flows occurring at the steady state to stabilise), from which the data on labour flows is retrieved and used to inform the calibration. 
The transition density matrices corresponding to labour flows observed in the steady state are denoted by $\mathbb{R}_*$, $\mathbb{I}_*$, and $\mathbb{O}_*$. 
Their counterparts generated from the observed LFNs are defined as $\mathcal{R}$, $\mathcal{I}$, and $\mathcal{O}$.

The steady state labour flows inform a multi-objective gradient descent algorithm (\autoref{alg:gdalg}) inspired by \cite{guerrero_how_2022}. 
Within each iteration of the algorithm we run a set of $M$ Monte Carlo simulations (i.e. independent simulations run with the same set of parameters). 
We define error matrices for each of the LFNs (region, industry, occupation) by $ e^R = \mathcal{R} - \overline{\mathbb{R}_*} $, $ e^I = \mathcal{I} - \overline{\mathbb{I}_*} $, and $ e^O = \mathcal{O} - \overline{\mathbb{O}_*} $ where, for example, $\overline{\mathbb{R}_*}$ indicates a matrix of flow density values averaged across the $M$ Monte Carlo simulations. 
The mean is representative of the underlying flows as, for a large enough agent population, flows will not vary substantially across the $M$ simulations.
Thus, we are aiming to minimise the difference between the observed flow densities (indicated within each cell of the flow density matrix) and their point estimates from the model.

We describe the parameter updating protocol taking the regional LFN as an example. 
For each cell of the error matrix $e^R$, if $e_{j,k}^R < 0$ then the density of flows between regions $j$ and $k$ in the simulation are higher than in the observed regional LFN. 
To reduce the magnitude of this error, we would like to decrease the density of flows between regions $j$ and $k$ in the simulation. 
We may be able to do so by making regions $j$ and $k$ less similar (from the agent's--subjective--point of view). 
This reduces the probability that an agent who currently holds a position in region $j$ (and is actively search for a new position) is matched with a position in region $k$.
It also decreases the probability that, if the agent from region $j$ is matched with and applies to a position in region $k$, they will subsequently be hired. 
For all pairs of regions where the geographical proximity value $R_{j,k}$ (calculated in \autoref{eqn:regionsimilarity}) satisfies $R_{j,k} \in (0,1)$, such a reduction can be achieved by multiplying $\nu_{j,k}^R$ by a factor $1 + \delta^R_{j,k}$. 
Here, $\delta^R_{j,k} = |e^R_{j,k}|/\mathcal{R}_{j,k}$ and is thus proportional to the magnitude of the error $e^R_{j,k}$. 
Increasing the value of $\nu^R_{j,k}$ reduces the value of $R_{j,k}^{\nu^R_{j,k}}$, making regions $j$ and $k$ less similar to each other.
For any $R_{j,k} \in \{0,1\}$, updating $\nu^R_{j,k}$ has no effect on the value of $R_{j,k}^{\nu^R_{j,k}}$. 
In these cases, we rely upon adjustments made to other cells of $\nu^R$ in response to the errors in the corresponding cells of $e^R$. 
A similar logic holds for the case where $e_{j,k}^R \geq 0$ and we wish to reduce the magnitude of the error by increasing the density of flows between regions $j$ and $k$.

\begin{figure}[h!]
	\centering
	\includegraphics[width=0.55\textwidth]{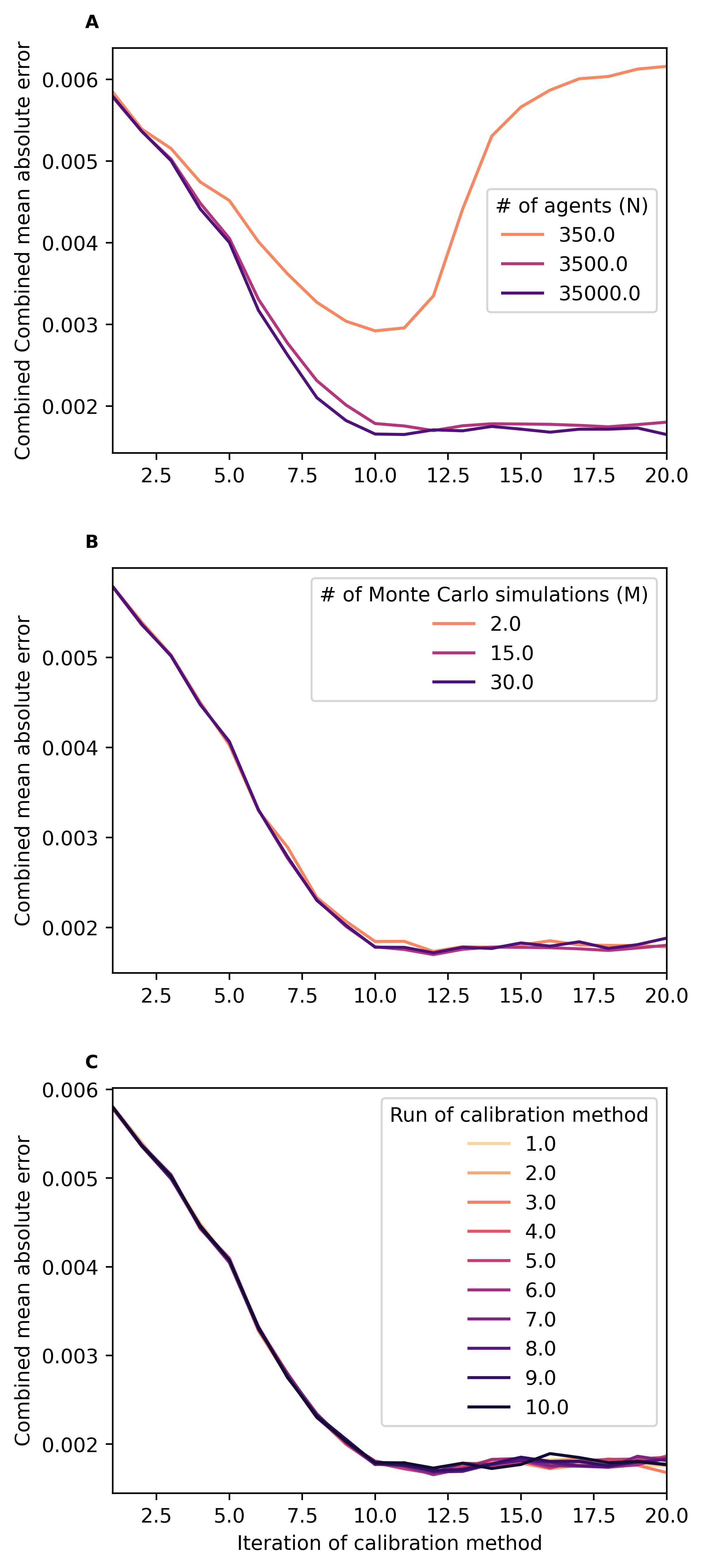}
	\caption{\textbf{Robustness of chosen calibration method.} The calibration method is robust to changes in a) the number of agents $N$ (except in the case of very small $N$) and b) the number of Monte Carlo simulations. Results also hold across multiple runs of the calibration method (with $N=3500$), as shown in c).}
	\label{figure:calibration_sensitivity}
\end{figure}

We track the behaviour of the combined mean absolute error (CMAE) metric, given by $ \text{CMAE} = \frac{1}{3} \left( \overline{|e^R|} + \overline{|e^I|} + \overline{|e^O|} \right)$, as we iterate over this error-reduction procedure, and select the threshold for this value such that the number of iterations is sufficient for the CMAE value to plateau at a minimum (\autoref{figure:calibration_sensitivity}). 
Similarly to \cite{guerrero_how_2022}, we bound our adjustment factor $(1 \pm \delta^\cdot)$ by 3/2 (respectively 1/2) as this substantially increases the rate at which we achieve error stabilisation. 
The calibration error is robust to the number of Monte Carlo simulations ($M$), and to the simulation scale (i.e. the number of agents, $N$) except in the case of a very low $N$-value (\autoref{figure:calibration_sensitivity}).
We note that, since we update all $\nu^R_{j,k}$ values simultaneously, we achieve good efficiency in comparison to a case where parameters need to be updated one-by-one. 
Additionally, the fact that we are able to directly estimate $\nu^R_{j,k}$ values is something that is often not possible in models with such a large number of parameters (where indirect inference and surrogate models are common approaches). 

\begin{algorithm}
    \SetKwInOut{Input}{Input}
    \SetKwInOut{Output}{Output}
    \text{Initialise $\nu^R, \nu^I, \nu^O$ as matrices filled with ones}\;
    \While{CMAE $<$ threshold}
      {
        \text{run $M$ Monte Carlo simulations}\;
        \text{compute error matrices $e^R, e^I, e^O$}\; 
        \ForEach{error matrix $e^\cdot$}
            {
                \eIf{$e^\cdot_{j,k} < 0 $}
                    {
                    $\nu^\cdot_{j,k} = \nu^\cdot_{j,k} \min \left(1 + \delta^\cdot, 3/2 \right)  $\;
                    }
                    {
                    $\nu^\cdot_{j,k} = \nu^\cdot_{j,k} \max \left(1 - \delta^\cdot, 1/2 \right) $\;
                    }
            }
      }
    \caption{Calibration pseudocode} \label{alg:gdalg}
\end{algorithm}

\section*{Shock experiments}

\begin{figure}[h!]
	\centering
	\includegraphics[width=0.95\textwidth]{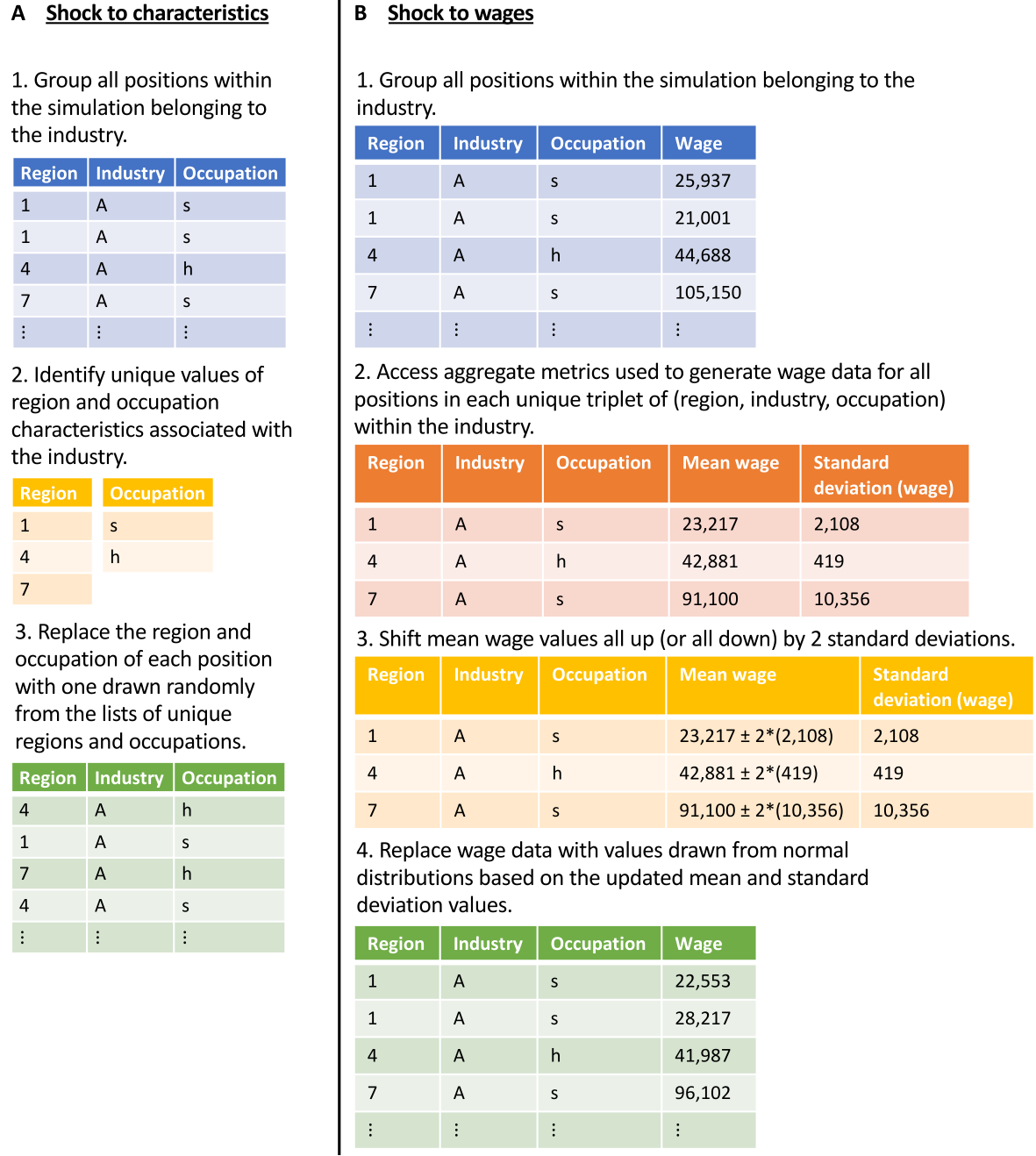}
	\caption{\textbf{Shock implementations.} Subplots indicate the method of implementing shocks that impact the a) characteristics and b) wages associated with all jobs within an industry.}
	\label{figure:shockschematic}
\end{figure}

Once the model has been calibrated, we run Monte Carlo simulations including a shock to the underlying job or wage distribution to determine how alterations to these inputs impact network structure. 
Shocks manifest as the homogenisation of characteristic(s) (i.e. region, occupation) associated with all positions in one or more industries, or changes to the wages associated with these positions (\autoref{figure:shockschematic}). 
It is possible to implement less generic shock scenarios (e.g. ones that specify a plausible change to the job and/or wage distribution driven by a specific shock, such as a pandemic). 
However, we utilise a systematic and stylised approach as we are seeking a more general understanding of how wages and job characteristics impact the topology of the LFN.
This avenue of inquiry motivates the need for models that emerge network structure.

For a shock impacting positions in $n$ industries the methods described in \autoref{figure:shockschematic} are applied to each industry individually. 
For example, we do not pool unique values for region and occupation across industries when applying a positional shock.
If a shock impacts wages, the actual change in mean wage will vary for positions within the industry according to their associated region and occupation.
This is because each set of positions (defined according to their region, occupation, and industry) has an associated mean and standard deviation for wage.

In all shock scenarios, the impact of the shock on the LFN structure is determined by comparing the edges in the unshocked network ($A$) to the edges in the shocked network ($B$). 
We compare the vectors of flow densities associated with these edges ($d^A$, $d^B$ respectively) using the weighted Jaccard distance, given by  

\begin{equation}
    1 - J_W(A,B) = 1 - \frac{\sum_{i=1}^{N} \min{\left(d^A_i, d^B_i \right)}}{\sum_{i=1}^{N} \max{\left(d^A_i, d^B_i \right)}}
\end{equation}
where $N$ is the total number of flows. 
This weighted metric allows us to move beyond a consideration of the presence/absence of flows (as would be captured by the unweighted Jaccard distance) to understand how densities are redistributed under a shock. 
We also examine how shocks impact the average clustering coefficient--which we interpret as the extent to which flows are localised--for shocked and unshocked networks. We limit ourselves to considering these features in order to focus on aspects of the network structure that have intuitive real-world interpretations.

Finally, we examine shock impacts at the level of individual flows (e.g. job-to-job transitions from region $i$ to $j$). 
We consider whether the distribution of differences in flow densities between shocked and unshocked networks differs significantly (according to a two-sided Mann-Whitney U test at a 5\% significance level) from the distribution of differences in flow densities between two sets of unshocked networks. 
In other words, do the flow densities under a simulation including a shock differ significantly from those under a simulation without a shock, if we account for the natural variation in flow density due to the stochastic nature of the model?
By focusing on flows where the density does differ significantly, we can understand how shocks re-distribute job-to-job transitions across the LFNs.

\bibliography{main}

\newpage

\title{Supplementary Information: Endogenous Labour Flow Networks}
\date{}

\maketitle

\setcounter{page}{1}
\counterwithin{figure}{section}
\counterwithin{table}{section}

\section*{Defining a steady state} \label{section:steadystatedefinition}

Let $\mathbb{R}_t$, $\mathbb{I}_t$, and $\mathbb{O}_t$ represent the model's regional, and occupational transition density matrices for the corresponding LFNs in period $t$.
The corresponding matrices for the observed LFNs are defined by $\mathcal{R}$, $\mathcal{I}$, and $\mathcal{O}$.
Thus, the error function for period $t$, given by $\Xi_t$, is defined as

\begin{equation}
    \Xi_t =  \frac{1}{3} \left( \lVert \mathbb{R}_t - \mathcal{R} \rVert_F  + \lVert \mathbb{I}_t - \mathcal{I} \rVert_F + \lVert \mathbb{O}_t - \mathcal{O}  \rVert_F \right),
    \label{eqn:calibrationerror}
\end{equation}
where $\lVert \cdot \rVert_F$ denotes the Frobenius norm of a matrix.

The steady state is reached if the following condition holds:
\begin{equation}
    \Bigg\lvert \left( \frac{1}{k} \sum_{t = T-k}^T  \Xi_t \right) -  \left(  \frac{1}{k} \sum_{t = T-k-l}^{T-l} \Xi_t \right) \Bigg\rvert < \epsilon
    \label{eqn:steadystatecondition}
\end{equation}
where $k$ is the size of the smoothing window, $l$ is the lag, $T$ is the current timestep of model, and $\epsilon$ is a small, positive threshold parameter.
In other words, the steady state is detected if there is stability and convergence in terms of the labour flows of the model.

\pagebreak

\section*{Dictionaries}

\begin{table}[h!]
\begin{center}
\begin{tabular}{| p{8cm} | p{4cm}|} 
 \hline
 Geographical Region & Label \\ [0.5ex] 
 \hline\hline
Tyne and Wear & Tyne \& Wear \\
 \hline
Rest of North East & R. of North East \\
 \hline
Greater Manchester & G. Manchester \\
 \hline
 Merseyside & Merseyside \\
 \hline
 Rest of North West & R. of North West \\
 \hline
 South Yorkshire & South Yorkshire \\
 \hline
 West Yorkshire & West Yorkshire \\
 \hline
 Rest of Yorkshire and Humberside & R. of Yorks. and H. \\
 \hline
 East Midlands & East Midlands \\
 \hline
 West Midlands (metropolitan county) & W. Midlands (MC) \\
 \hline
Rest of West Midlands & R. of W. Midlands \\
 \hline
 East of England & East of England \\
 \hline
 Central London & Central London \\
 \hline
 Inner London & Inner London \\
 \hline
 Outer London & Outer London \\
 \hline
 South East & South East \\
 \hline
 South West & South West \\
 \hline
 Wales & Wales \\
 \hline
 Strathclyde & Strathclyde \\
 \hline
 Rest of Scotland & R. of Scotland \\
 \hline
 Northern Ireland & N. Ireland \\
 \hline
\end{tabular}
\end{center}
\caption{\textbf{Region dictionary.} Labels used to refer to geographical regions within the UK throughout text.}
\label{tab:region codes}
\end{table}

\begin{table}[h!]
\begin{center}
\begin{tabular}{| p{10cm} | p{3cm}|} 
 \hline
 Industry section & Label \\ [0.5ex] 
 \hline\hline
Agriculture, forestry and fishing & Agriculture \\
 \hline
 Mining and quarrying & Mining \\
 \hline
 Manufacturing &	Manufacturing \\
 \hline
Electricity, gas, steam and air conditioning supply	& Electricity \\
 \hline
Water supply, sewerage, waste management and remediation activities	& Water \\
 \hline
Construction	& Construction \\
 \hline
Wholesale and retail trade, repair of motor vehicles and motorcycles	& Motor trade \\
 \hline
Transportation and storage	& Transport \\
 \hline
Accommodation and food service activities	& Accommodation \\
 \hline
Information and communication	& Infocomm \\
 \hline
Financial and insurance activities	& Finance \\
 \hline
Real estate activities	& Real estate \\
 \hline
Professional, scientific and technical activities	& Science \& tech \\
 \hline
Administrative and support service activities	& Admin \\
 \hline
Public administration and defence, compulsory social security	& Public \\
 \hline
Education	& Education \\
 \hline
Human health and social work activities & Health \& social \\
 \hline
 Arts, entertainment and recreation & Entertainment \\
 \hline
 Other service activities & Other service \\
 \hline
Activities of households as employers; undifferentiated goods- and services-producing activities of households for own use & Household \\
 \hline
 Activities of extraterritorial organisations and bodies & Extra orgs \\
 \hline
\end{tabular}
\end{center}
\caption{\textbf{Industry dictionary.} Labels used to refer to SIC industry sections throughout text.}
\label{tab:industry codes}
\end{table}

\begin{table}[h!]
\begin{center}
\begin{tabular}{| p{9cm} | p{2.5cm}|} 
 \hline
Occupation major group & Label \\ [0.5ex] 
 \hline\hline
Managers, directors and senior officials	& Manager \\
 \hline
Professional occupations	& Professional \\
 \hline
Associate professional occupations	& Ass. prof. \\
 \hline
Administrative and secretarial occupations	& Admin \\
 \hline
Skilled trades occupations	& Trades \\
 \hline
Caring, leisure and other service occupations	& Leisure \\
 \hline
Sales and customer service occupations	& Sales \\
 \hline
Process, plant and machine operatives	& Machine op. \\
 \hline
Elementary occupations	& Elementary \\
 \hline
\end{tabular}
\end{center}
\caption{\textbf{Occupation dictionary.} Labels used to refer to SOC major groups throughout text.}
\label{tab:occupation codes}
\end{table}

\clearpage

\section*{Supplementary figures}

\section*{Data availability}

\noindent All data required to run simulations that can be made publicly available are located in a GitHub repository (\url{https://github.com/alan-turing-institute/UK-LabourFlowNetwork-Model}). Data generated from the LFS cannot be made publicly available under the ONS statistical data disclosure rules.

\subsection*{Data usage statements}

This project includes information from \href{https://www.onetcenter.org/}{O*NET 26.2 Database} by the U.S. Department of Labor, Employment and Training Administration (USDOL/ETA). Used under the \href{https://creativecommons.org/licenses/by/4.0/}{CC BY 4.0} license. O*NET \textsuperscript{\textregistered} is a trademark of USDOL/ETA.

We also use data from OpenStreetMap (\textsuperscript{\textcopyright} OpenStreetMap contributors) licensed under the \href{https://opendatacommons.org/licenses/odbl/}{Open Data Commons Open Database License}.

Skills data powered by \href{https://www.lmiforall.org.uk/}{LMI for All}. 

\section*{Code availability}
\noindent Code used for data analysis, parameter fitting, and simulation is available in a GitHub repository (\url{https://github.com/alan-turing-institute/UK-LabourFlowNetwork-Model}).

\section*{Acknowledgements}

\noindent This work was supported by Wave 1 of The UKRI Strategic Priorities Fund under the EPSRC Grant EP/W006022/1, particularly the ``Shocks and Resilience" theme within that grant \& The Alan Turing Institute. We would like to thank Áron Pap for conducting preliminary explorations of this topic, and Andy Jones and his team at BEIS for their interactions with us throughout this project. We also thank Dr. Alden Conner for her contributions to the development of the associated code repository.

\section*{Author information}

\subsection*{Contributions}
\noindent K.R. Fair (Methodology, Software, Validation, Formal analysis, Investigation, Data curation, Writing - Original Draft, Writing - Review \& Editing, Visualisation). 

O.A. Guerrero (Conceptualization, Methodology, Software, Writing - Original Draft, Writing - Review \& Editing, Supervision, Project administration).

\subsection*{Correspondence}

\noindent Correspondence and material requests should be addressed to Kathyrn R Fair.

\section*{Competing interests}
\noindent The authors declare no competing interests.

\end{document}